\documentclass[%
reprint,
 amsmath,amssymb,
 aps,
]{revtex4-2}
\usepackage{color}
\usepackage{graphicx}
\usepackage{dcolumn}
\usepackage{bm}
\usepackage{amsmath,amsbsy,amssymb}
\usepackage{mathrsfs}
\usepackage{ulem}
\usepackage{mathcomp}
\usepackage{physics}
\usepackage{comment}
\usepackage{ascmac}
\usepackage{here}

\newcommand{\la}{\langle}
\newcommand{\ra}{\rangle}
\newcommand{\al}{\alpha}
\newcommand{\sg}{\sigma}
\newcommand{\gm}{\gamma}

\begin{document}

\preprint{APS/123-QED}

\title{Gate-tunable giant superconducting nonreciprocal transport in few-layer $T_d$-MoTe$_2$}

\author{T. Wakamura,$^1$ M. Hashisaka,$^{1,2}$ S. Hoshino,$^3$, M. Bard$^1$, S. Okazaki,$^4$ T. Sasagawa,$^4$ T. Taniguchi,$^5$ K. Watanabe,$^6$ K. Muraki,$^1$}
\author{N. Kumada$^1$}
\affiliation{$^1$NTT Basic Research Laboratories, NTT Corporation, 3-1 Morinosato-Wakamiya, Atsugi, 243-0198, Japan}
\affiliation{$^2$Institute for Solid State Physics,University of Tokyo, 5-1-5 Kashiwa-no-ha, Kashiwa 277-8581, Japan}
\affiliation{$^3$Department of Physics, Saitama University, Shimo-Okubo, Saitama 338-8570, Japan}
\affiliation{$^4$Laboratory for Materials and Structures, Tokyo Institute of Technology, Nagatsuta, 226-8503, Japan}


\affiliation{$^5$International Center for Materials Nanoarchitectronics, National Institute for Materials and Science, 1-1 Namiki, Tsukuba, 305-0044, Japan}
\affiliation{$^6$Research Center for Functional Materials, National Institute for Materials and Science, 1-1 Namiki, Tsukuba, 305-0044, Japan}


\date{\today}

\begin{abstract}
 We demonstrate gate-tunable giant field-dependent nonreciprocal transport (magnetochiral anisotropy) in a noncentrosymmetric superconductor $T_{\rm d}$-MoTe$_2$ in the thin limit. Giant magnetochiral anisotropy (MCA) with a rectification coefficient (or a figure of merit) $\gamma$ = $3.1 \times 10^6$ T$^{-1}$ A$^{-1}$ is observed at 230 mK, below the superconducting transition temperature ($T_c$). This is one of the largest values reported so far and may be attributed to the reduced symmetry of the crystal structure. The temperature dependence of $\gamma$ indicates that ratchet-like motion of magnetic vortices is the origin of the MCA, as supported by our theoretical model. For bilayer (2 L) $T_{\rm d}$-MoTe$_2$, \textcolor{black}{we can successfully modulate $\gamma$ by gating}. Our experimental results provide a new route to realizing electrically controllable superconducting rectification devices in a single material.  
\end{abstract}

\maketitle



Recent intensive studies on nonreciprocal transport have revealed the potential of using noncentrosymmetric materials or inversion-symmetry-breaking multilayer structures to develop novel rectification devices \textcolor{black}{based on superconducting or Josephson diode effect} \cite{SCDiode_Ono, SCDiode_Ali, SCDiode_NbSe2}. In systems with broken inversion and time-reversal symmetries, Onsager's reciprocal theorem allows the electrical resistance to be different for opposite current directions. This is called magnetochiral anisotropy (MCA), which leads to the rectification effect \cite{Tokura, Ideue, Ando}.

Broken inversion symmetry is more beneficial in superconductors. Rectification via ratchet-like motion of magnetic vortices was reported more than a decade ago for superconductors with asymmetric artificial magnetic nanostructures or with asymmetric antidots as an asymmetric pinning potential \cite{Ratchet1, Ratchet2, Ratchet_PRB1, Ratchet_PRB2, Zhu, Villegas}. These previous works revealed that as the asymmetry of the pinning potential for magnetic vortices becomes stronger, rectification becomes more efficient. Recent studies have pointed out that the ratchet-like motion of magnetic vortices is also possible in unpatterned noncentrosymmetric superconductors and provides large MCA \cite{Hoshino, Itahashi, Ideue2, Wakatsuki}. In such systems, the asymmetry of the crystal structure intrinsically induces asymmetric pinning potential. In contrast to superconducting films with artificial structures, noncentrosymmetric superconductors do not require complex fabrication processes, offering more facile accessibility to nonreciprocal transports. However, previous reports on MCA in noncentrosymmetric superconductors has mostly concentrated on those with trigonal symmetry \cite{MoS2, NbSe2, Itahashi, Ideue2, Itahashi2}, and other crystal symmetries have been poorly investigated. Toward more efficient rectification via MCA, it is essential to explore noncentrosymmetric superconductors with different symmetries, especially with lower crystal symmetry than trigonal symmetry. Furthermore, regarding future technological applications, facile electrical control of nonreciprocal signals is required, but no previous reports have addressed gate modulation of superconducting MCA.

\begin{figure}[tb!]
\begin{center}
\includegraphics[width=8cm,clip]{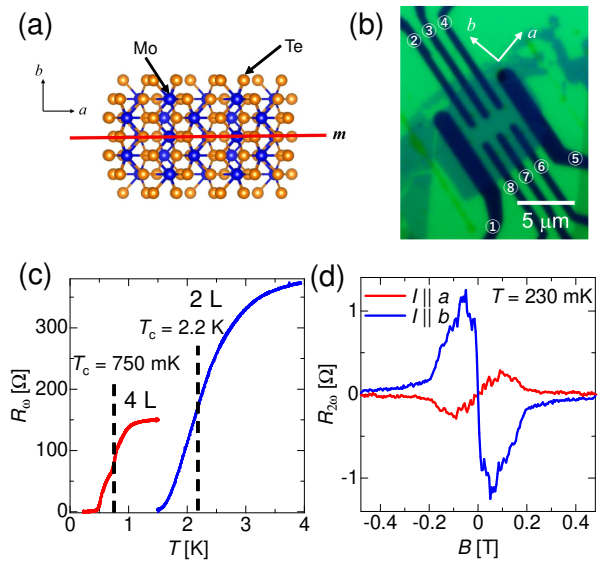}
\caption{(a) Top view of the crystal structure, in which broken inversion symmetry is evident. For thin layers, only one mirror plane is present. (b) Optical microscope image of a 4 L device. A thin $T_{\rm d}$-MoTe$_2$ flake is deposited on metallic contacts prepared in advance. (c) Temperature dependence of the resistance of 4 L and 2 L samples. (d) Comparison of $R_{2\omega}$ when current is parallel to the $a$-axis (red) and $b$-axis (blue) taken of the 4 L sample. For the setup with $I \parallel a$, we drive the current between \textcircled{\scriptsize 1} and \textcircled{\scriptsize 5} and measure the voltage between \textcircled{\scriptsize 3} and \textcircled{\scriptsize 4}. For the $I \parallel b$ configuration, the current is driven between \textcircled{\scriptsize 3} and \textcircled{\scriptsize 7}, and the voltage is measured between \textcircled{\scriptsize 2} and \textcircled{\scriptsize 8} (see also Figs.~\ref{current_dist}).}
\label{fig1}
\end{center}
\end{figure}

In this study, we demonstrate gate-tunable giant MCA in a noncentrosymmetric superconductor $T_{\rm d}$-MoTe$_2$ in the thin limit. $T_{\rm d}$-MoTe$_2$ lacks inversion symmetry and, for thin layers, has only one mirror plane normal to the $b$-axis as shown in Fig.~\ref{fig1}(a). This reduced symmetry of the crystal structure may make the pinning potential for magnetic vortices highly asymmetric, which can contribute to generating large MCA. From the MCA measurements under a perpendicular magnetic field in few-layer $T_{\rm d}$-MoTe$_2$ samples below $T_c$, we obtain 
the rectification coefficient, i.e. a figure of merit $\gamma$ = 3.1 $\times$ 10$^6$ T$^{-1}$A$^{-1}$ at 230 mK, one of the largest values among those reported so far. The monotonic increase in $\gamma$ with decreasing temperature indicates that the giant MCA is due to the ratchet-like motion of magnetic vortices in the mixed state of the type-II superconductor \cite{Hoshino}. Interestingly, despite that  $T_{\rm d}$-MoTe$_2$ is a semimetal, in the 2 L sample we can successfully modulate the MCA via an external gate voltage and demonstrate the modulation of $\gamma$. This ability to produce a large variation in the nonreciprocal resistance by changing the gate voltage may provide key insights into the mechanisms behind the giant MCA by associating it with modulation of the superconducting properties.  

The few-layer $T_{\rm d}$-MoTe$_2$ flakes are mechanically exfoliated from high-quality $T_{\rm d}$-MoTe$_2$ crystals with a residual-resistivity ratio (RRR) $\sim$ 1000 grown via the flux growth method. The mechanical exfoliation is carried out inside an Ar-filled glovebox containing concentrations of O$_2$ and H$_2$O below 0.5 ppm. Independently from the flakes, we prepare metallic electrodes by using typical electron-beam (EB) lithography and EB evaporation on a SiO$_2$/Si substrate (for the 4 L sample) or hexagonal boron-nitride (h-BN) exfoliated on a SiO$_2$/Si substrate (for the 2 L sample). We use Au (for the 4 L sample) or Pt (for the 2 L sample) with Ti as a buffer layer for the electrodes, and the total thickness of the electrodes is set to less than 10 nm to avoid exerting extra strain on the flake. The exfoliated $T_{\rm d}$-MoTe$_2$ is transferred with h-BN picked up by using poli-dimethylpolysiloxane (PDMS) covered with polycarbonate (PC) film. The h-BN flake on $T_{\rm d}$-MoTe$_2$ also acts as a protective layer because $T_{\rm d}$-MoTe$_2$ easily deteriorates when it is exposed to air. The transfer process is also performed inside the glovebox. After taking the sample out of the glovebox, it is immediately mounted on the sample holder of the $^3$He insert and encapsulated by the internal vacuum can then cooled down. Raman scattering and atomic-force microscope (AFM) measurements are carried out at room temperature and in the atmosphere after the transport measurements. 


In materials under broken inversion and time-reversal symmetries, Onsager's reciprocal theorem allows the linear longitudinal resistance to be different for opposite current directions \cite{Tokura}. Rikken \textit{et al}. heuristically found a general formula for the nonreciprocal transport, also called the MCA, expressed as \cite{Rikken}
\begin{equation}
R = R_0(1 + \gamma BI),
\label{eq1}
\end{equation}
where $\gamma$ is the rectification coefficient, which quantifies the efficiency of generating the nonreciprocal resistance. $R_0$, $B$ and $I$ are the linear resistance, magnetic field, and excitation current, respectively. Substituting equation (\ref{eq1}) into Ohm's law $V = RI$ leads to
\begin{equation}
V = R_0 I+ \gamma BR_0I^2.
\label{eq2}
\end{equation}
The first term is the typical linear voltage response to the current and the second term is related to the nonreciprocal transport. Thus, the nonreciprocal response is obtained as a second harmonic signal for the ac excitation current $I_\omega \propto \sin(\omega t)$.

\begin{figure*}[tb]
\begin{center}
\includegraphics[width=16cm,clip]{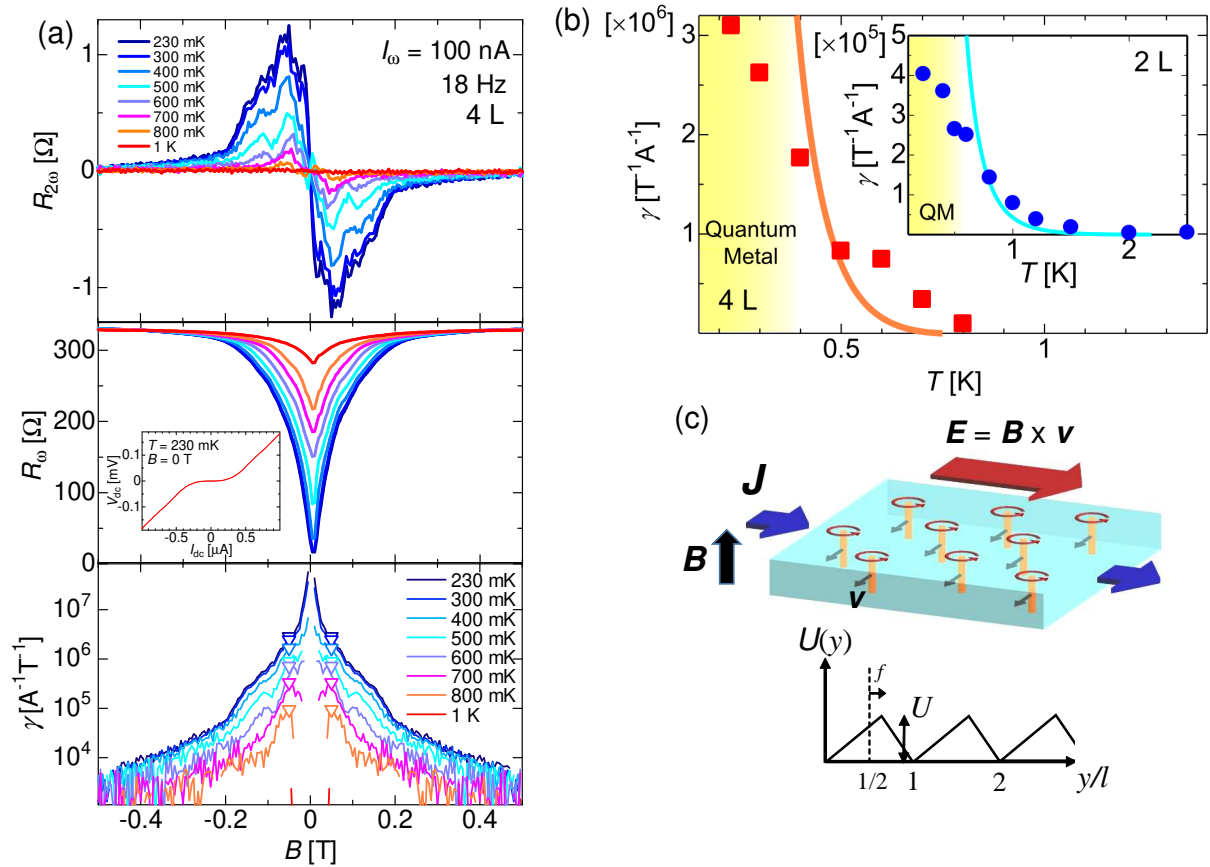}
\caption{(a) Experimental data from the 4 L sample. ~Top: Nonreciprocal resistance ($R_{\rm 2 \omega}$ = $V_{\rm 2 \omega}/I_{\rm \omega}$, $I_\omega$ = 100 nA) measured at different temperatures. Middle: $R_\omega$ signals measured simultaneously with $R_{2 \omega}$. \textcolor{black}{Bottom: $\gamma$ as a function of $B$ at different temperatures. The $B$ values where $R_{2 \omega}$ shows a peak ($B_{\rm peak}$) are marked with open triangles.} \textcolor{black}{Inset in the middle figure: $I_\omega-V_\omega$ curve at $B$ = 0 T and $T$ = 230 mK. Zero resistance is observed around $I_\omega$ = 0, but the shape of the curve is somewhat rounded, partly because of the 2D nature of superconductivity.} 
(b) Temperature dependence of $\gamma$ taken from the 4 L sample. The orange curve shows the fit based on equation (\ref{eq3}). Inset: Experimental data of $\gamma$ as a function of temperature obtained from the 2 L sample with the fit. \textcolor{black}{Yellow shaded regions in the main figure and the inset represent the quantum metal (QM) phase (see also Appendix E).}
(c) Top: Schematic illustration of the motion of magnetic vortices with the velocity $\mathbf{v}$ driven by an external current $\mathbf{J}$, which generates an electric field $\mathbf{E} = \mathbf{B} \times \mathbf{v}$. Bottom: Image of a sawtooth potential assumed as the ratchet pinning potential in (\ref{eq3}).}
\label{fig2}
\end{center}
\end{figure*} 

First, we show the temperature dependence of the resistance for the four layer [4 L, see Fig.~\ref{fig1}(b)] and bilayer (2 L) samples in Fig.~\ref{fig1}(c). While $T_c$ is low ($\sim$ 100 mK) for bulk $T_{\rm d}$-MoTe$_2$ \cite{Qi}, that for the 4 L and 2 L samples is 750 mK and 2.2 K, respectively. This large enhancement in $T_c$ for thin layers is consistent with previous studies \cite{Rhodes, MoTe2cn}. Note that here $T_c$ is defined as the temperature where the resistance becomes half of that in the normal state. 


Now let us focus on measuring the nonreciprocal transport in the superconducting state. Figure~\ref{fig1}(d) shows the second-harmonic longitudinal resistance $R_{2\omega}$ for $I_\omega \parallel b$ and for $I_\omega \parallel a$ at 230 mK. $a$ and $b$ are the crystal axes as defined in Fig.~\ref{fig1}(a), and the $b$-axis is orthogonal to the mirror plane. A clear peak and dip are observed in $R_{2\omega}$ for $I_\omega \parallel b$. The field-asymmetric $R_{2 \omega}$ signals are in agreement with the MCA in (\ref{eq1}) and are consistent with previous experimental results \cite{MoS2, NbSe2, SrTiO3, Itahashi}. \textcolor{black}{The suppression of $R_{2 \omega}$ at higher fields is due to the suppression of superconductivity by a magnetic field.} Note that the nonlinearity of the resistance due to the transition between the normal and superconducting state is symmetric in $B$, so it is excluded as the origin of $R_{2 \omega}$. In contrast to the case for $I_\omega \parallel b$, $R_{2 \omega}$ for $I_\omega \parallel a$ is dramatically suppressed. This is also consistent with the geometry of MCA, where the symmetry plane, the directions of the magnetic field, and generated second-harmonic voltage are all perpendicular to each other \cite{Tokura}. Note that the finite signal for $I_\omega \parallel a$ is due to misalignment of the electrodes to the crystal axis [see Fig.~\ref{fig1}(b) and also Appendices B, F and G]. Below we focus on the geometry where $I_\omega \parallel b$.

The top part of Fig.~\ref{fig2}(a) displays $R_{2 \omega}$ as a function of perpendicular magnetic field measured at different temperatures. The amplitude of the signals monotonically decreases with increasing temperature. Above $T_c$, $R_{2 \omega}$ is completely suppressed, indicating that the effect is related to superconductivity. The middle part of Fig.~\ref{fig2}(a) shows the $R_\omega$ signals measured simultaneously with the $R_{2 \omega}$ signals. \textcolor{black}{The inset is $I_\omega-V_\omega$ curve measured at $B$ = 0 T and $T$ = 230 mK, showing the exactly zero resistance state at low $I_\omega$. At $I_\omega$ = 100 nA, $R_\omega$ starts to deviate from zero immediately after a magnetic field is turned on.}

Now that we have obtained $R_{2 \omega}$ and $R_{\omega}$, we can estimate the value of the rectification coefficient $\gamma = 2 R_{2 \omega}/(R_\omega B I_\omega)$ \cite{Tokura, MoS2, Hoshino}. \textcolor{black}{As shown in the bottom of Fig.~\ref{fig2}(a), $\gamma$ depends on $B$ and increases rapidly as $B \rightarrow$ 0, particularly at the lowest temperature. This is due to the decrease in $R_\omega$ and $B$ in the denominator. In the following, we use $\gamma$ at $B$ ($\equiv B_{\rm peak}$), at which $R_{2 \omega}$ is at a peak [triangles in the bottom of Fig.~\ref{fig2}(a)] as a representative value. This definition of representative $\gamma$ is often used in previous studies and useful for quantitative discussion. \cite{Masuko, MoS2, NbSe2, Itahashi, Ideue2}.} Figure~\ref{fig2}(b) shows that $\gamma$ continues to increase with decreasing temperature and reaches $\gamma$ = 3.1 $\times$ 10$^6$ T$^{-1}$ A$^{-1}$ at 230 mK, the lowest measurement temperature. This value is two to three orders of magnitude larger than that of other two-dimensional superconductors, such as MoS$_2$ and NbSe$_2$ as we will discuss later. In the inset of Fig.~\ref{fig2}(b), we also plot the temperature dependence of $\gamma$ for the 2 L sample, which shows the similar trend with slightly smaller amplitudes. \textcolor{black}{The smaller $\gamma$ in the 2 L sample is due to larger $B_{\rm peak}$ caused by more robust superconductivity (see Appendix D)}. 

\begin{table*}[tb!]
\caption{\label{tab:table3}Summary of $\gamma$, \textcolor{black}{$\gamma W$, $\gamma Wt$} and $\gamma B_{\rm peak}$ for different 2D superconductors}
\begin{ruledtabular}
\begin{tabular}{cccccccccc}
 Material&Symmetry&$B_{\rm peak}$ [T] & $R_{2 \omega} [\Omega]$ & $W$ [$\mu$m] & $\gamma$ [A$^{-1}$T$^{-1}$] & \textcolor{black}{$\gamma W$ [A$^{-1}$T$^{-1}$m]} & \textcolor{black}{$\gamma Wt$ [A$^{-1}$T$^{-1}$m$^2$]} &$\gamma B_{\rm peak}$ [A$^{-1}$] & ref.\\ \hline
 MoTe$_2$ 2 L & $C_{s}$ & 0.35 & 1.8 & 3 & 6.7 $\times 10^5$ & 2.0 & 2.8 $\times 10^{-9}$ & 2.34 $\times 10^5$ & This study \\
 MoTe$_2$ 4 L & $C_{s}$ & 0.050 & 1.3 & 6 & 3.1 $\times 10^6$ & 19 & 1.1 $\times 10^{-7}$ & 1.54 $\times 10^5$ & This study \\
MoS$_2$\footnotemark[1]& $C_{3v}$ & 0.50 & 0.65 & 3 & 4.6$\times10^3$ & 1.4 $\times 10^{-2}$ & - & 2.32$\times10^3$ & \cite{Wakatsuki}\\
 NbSe$_2$ 5 L& $D_{3h}\footnotemark[2]$ & 3.0 & 0.062 & 5 & 2.8$\times10^2$ & 1.4$\times 10^{-3}$ &4.7 $\times 10^{-11}$ & 8.49$\times10^2$ & \cite{NbSe2} \\
 SrTiO$_3$\footnotemark[1]& - & 0.050 & 0.29 & 80 & 3.2 $\times 10^6$ & 2.6 $\times 10^2$ &-& 1.60 $\times 10^5$ & \cite{Itahashi} \\
\end{tabular}
\end{ruledtabular}
\footnotetext[1]{\textcolor{black}{Because superconductivity is induced by the ionic liquid gating close to the surface, the exact value of $t$ is difficult to define.}}
\footnotetext[2]{The point group of bulk NbSe$_2$ is $D_{6h}$. On the other hand, $D_{3h}$ is the point group for thin NbSe$_2$ with the odd number of layers.}
\end{table*}

So far, several mechanisms have been proposed to explain the MCA in the superconducting state \cite{Wakatsuki, Hoshino}. The temperature dependence of the signals and the direction of the applied magnetic field are clues for identifying the mechanism. For example, paraconductivity is a mechanism proposed as an origin of MCA under an in-plane magnetic field \cite{Hoshino,SrTiO3}. Since it is relevant to thermal fluctuations of the superconducting order parameter, the nonreciprocal signal is slightly enhanced above $T_c$ and suppressed much below $T_c$. On the other hand, ratchet-like motion of magnetic vortices enhances the MCA below $T_c$ under a perpendicular magnetic field \cite{Hoshino}. In the mixed state of type-II superconductors, magnetic fluxes penetrate the superconductor, and they are usually trapped by pinning potentials induced by disorder \textcolor{black}{such as underlying discrete lattice structure, defects, and impurities}. External current can drive the magnetic fluxes through the Lorenz force as schematically shown in the top image of Fig.~\ref{fig2}(c), if it is large enough to overcome the pinning potential \cite{Tinkham2, Vortex}. In superconductors with broken inversion symmetry, the asymmetry of the crystal structure locally affects the shape of the pinning potentials, making them asymmetric \cite{Rikken, Fente}. In this case, the magnetic vortices can exhibit ratchet-like motion, where the leftward and rightward motion of the vortex is not equivalent \cite{Hoshino, Ideue2, Itahashi, NbSe2, Ratchet1, Ratchet2, Zhu, Ratchet_PRB1, Ratchet_PRB2}. This asymmetry provides a source for nonreciprocal transport. 
The ratchet-like motion of magnetic vortices provides increasing $\gamma$ with decreasing temperature because thermal fluctuations of the magnetic vortices inside the pinning potential, which disturb the ratchet-like motion, are suppressed with decreasing temperature, and also the coherence length, which determines the diameter of the vortex, becomes smaller, making the vortex more sensitive to the pinning potentials.

\begin{figure*}[tb]
\begin{center}
\includegraphics[width=16.5cm,clip]{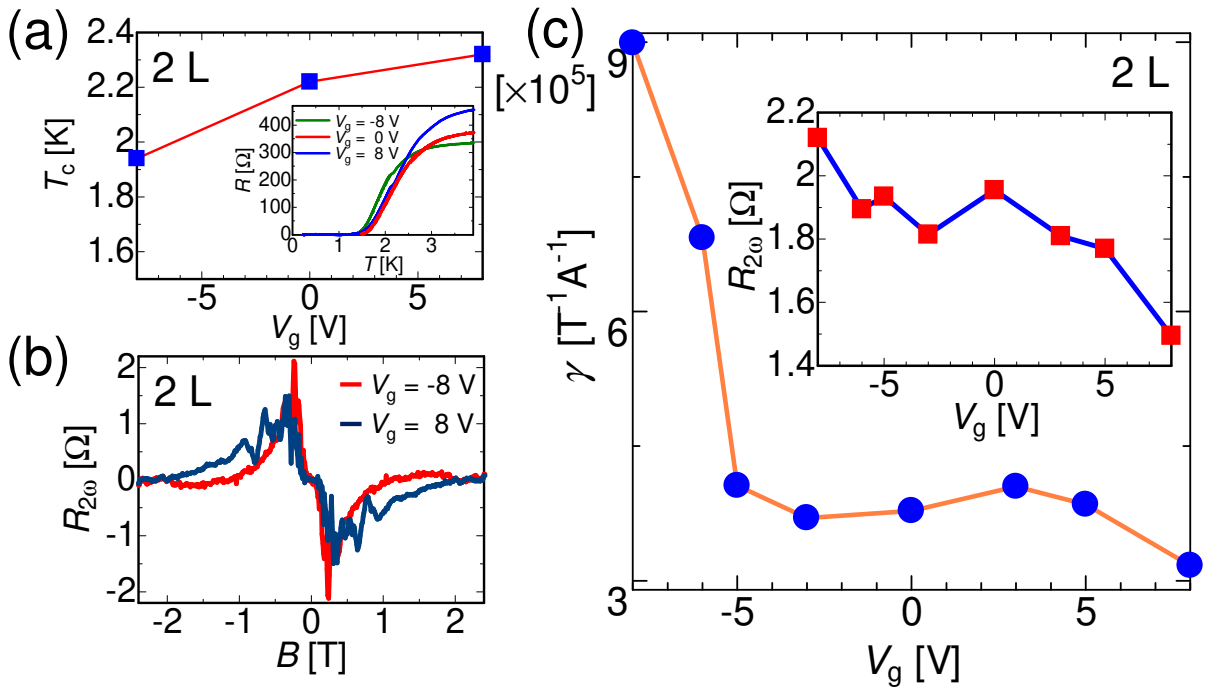}
\caption{(a) Gate voltage ($V_g$) dependence of $T_c$ for the 2 L sample taken at 230 mK. The inset shows the $R$-$T$ curves for the different $V_g$. (b) $R_{2\omega}$ as a function of $B$ at different $V_g$ at 230 mK. (c) $\gamma$ as a function of $V_g$. Inset: Gate voltage dependence of $R_{2 \omega}$.}
\label{fig4}
\end{center}
\end{figure*} 

\textcolor{black}{We next discuss the temperature dependence of $\gamma$ in more detail by comparing the experimental data with a theoretical model based on the ratchet-like motion of magnetic vortices. Solid lines in Fig.~\ref{fig2}(b) represent the theoretical curve following the expression (see Appendix H for details) \cite{Hoshino, Itahashi2}}:
\begin{equation}
\gamma = \frac{\phi_0^* \beta \ell } {W B}
\, 
\frac{g_2(\beta U)}{g_1(\beta U)}
, \label{eq3}
\end{equation}
where $W$ is the width of the sample, $\phi_0^*=h/2|e|$ is the flux quantum and $\beta = 1/k_B T$ is the inverse temperature.
$\ell$ and $U$ are the mean periodicity and the height of the pinning potential for a vortex, respectively. \textcolor{black}{Here, in order to account for vortex dynamics qualitatively, we phenomenologically introduce the pinning potential and employ the Langevin dynamics.}
We take the simple potential shape shown in the bottom figure of Fig.~\ref{fig2}(c), where
the dimensionless parameter $f$ controls the asymmetry of the potential. 
$g_{1}$ and $g_2$ are dimensionless functions determined from the linear- and second-order responses. 
The ratio is given by $\frac{g_2(\beta U)}{g_1(\beta U)} \sim \frac{f(\beta U)^3}{180}$ for a moving vortex regime with a small ratchet potential.
The curves follow the experimental data qualitatively in the intermediate temperature region as shown in Fig.~\ref{fig2}(b),
which supports the ratchet-like motion of magnetic vortices as the dominant mechanism for the giant MCA in this system. 
In the theoretical curves, there are two fitting parameters $\alpha$ and $f$, and the former defines the exponent in the temperature dependence of the ratchet potential $U \sim U_0 [(T_c - T)/T_c]^\alpha$ with $U_0$ as $U(T = 0)$. Fitting the experimental results in the intermediate temperature region provides $\alpha$ and $f$. By using these values and $U$ estimated from the critical current density $j_c$ at a small magnetic field, we obtain $U_0$ = 0.10 (0.17) meV for the 4 L (2 L) sample. Note that these $U_0$ values are consistent with $U_0$ = 0.10 (0.56) meV, alternatively obtained from the temperature dependence of the resistance under different magnetic fields (see Appendix E).  

\textcolor{black}{In contrast to the intermediate temperature region, the fits substantially overestimate $\gamma$ at lower temperatures. This can be explained by appearance of the quantum metal phase, in which quantum tunneling of vortices through the pinning potential suppresses the ratchet-like motion and thus MCA \cite{Itahashi, Hamamoto}. Indeed, the temperature range where the experimental data deviate from the theoretical curve corresponds to the region for the quantum metal phase [yellow shaded region on Fig. \ref{fig2}(b)] in the vortex phase diagram (see Appendix E). Note that the current density in the present measurements is small enough to preserve the quantum metal phase \cite{Itahashi}.
At higher temperatures, on the other hand, rectification mechanism is replaced by superconducting fluctuation and normal contributions \cite{Hoshino,SrTiO3}.}

We then compare the amplitude of $\gamma$ with those in different materials. The summary of $\gamma$ in different noncentrosymmetric superconductors is shown in Table I. $T_{\rm d}$-MoTe$_2$ exhibits larger $\gamma$ by several orders of magnitude than those for trigonal superconductors such as MoS$_2$ and NbSe$_2$. The only value comparable to ours reported in the previous studies is that from a SrTiO$_3$ Rashba superconductor under an in-plane magnetic field \cite{SrTiO3}. 
Therefore, the value obtained in our study is one of the largest reported so far \cite{MoS2, NbSe2, Itahashi, Ideue2, Itahashi2}. 

In Table I, we also show other quantities taking into account the difference in the sample geometry and also $B_{\rm peak}$. Because $\gamma$ depends on the sample width \textcolor{black}{and thickness}, we compare $\gamma W$ and $\gamma W\textcolor{black}{t}$, where $W$ \textcolor{black}{and $t$} are the width \textcolor{black}{and the thickness} of the sample, respectively. It is evident that the values for $T_{\rm d}$-MoTe$_2$ surpasses substantially those for other noncentrosymmetric superconductors, \textcolor{black}{except for SrTiO$_3$ due to the much larger sample width.}
The values of $\gamma B_{\rm peak}$ are also compared, which consider the difference in $B_{\rm peak}$. $T_{\rm d}$-MoTe$_2$ shows much larger values than those for MoS$_2$ and NbSe$_2$, and is comparable or slightly larger than the value for SrTiO$_3$. These comparisons corroborate the enhanced MCA in $T_{\rm d}$-MoTe$_2$ among van der Waals superconductors. Note that while our lowest measurement temperature ($\equiv T_{\rm meas}$) is lower than that in previous studies for MoS$_2$ or NbSe$_2$ \cite{MoS2, NbSe2, Itahashi}, $T_c$ of $T_{\rm d}$-MoTe$_2$ is lower, and the energy scale ($k_B T_{\rm meas})$ relative to the superconducting gap is comparable. Thus we can rule out lower measurement temperature as an origin for the substantial nonreciprocal signals. 

Difference in the crystal symmetry is likely to be the origin of the large variation in $\gamma$. In comparison with other two-dimensional trigonal superconductors, $T_{\rm d}$-MoTe$_2$ has reduced symmetry with only one mirror plane for thin layers. This reduced symmetry affects the asymmetry of the pinning potential. Since the symmetry of the pinning potential is crucial for the vortex dynamics, as reported previously \cite{Zhu, Villegas, Palau, Morgan}, the lower symmetry in the pinning potentials may generate larger nonreciprocal signals. Further theoretical study is required to scrutinize the effect of the symmetry reduction on the amplitude of nonreciprocal signals.


Finally, we demonstrate the gate modulation of the MCA for the 2 L $T_{\rm d}$-MoTe$_2$. While gate control of the MCA in the normal state has been studied in a BST topological nanowire \cite{Ando} and at the LaAlO$_3$/SrTiO$_3$ interface \cite{LAO}, it has not been reported yet in superconductors. The primary reason is that the concentration of charge carriers in a superconductor is typically high, making it challenging to employ a conventional solid gate to regulate superconducting characteristics due to the electric field screening on the nanometer scale within the material. We can overcome this problem by thinning down $T_{\rm d}$-MoTe$_2$ to a thickness comparable to the screening length \cite{Ma, Ferro}. \textcolor{black}{Note that the screening length is estimated to be $\sim$ 0.4 nm, the same order of the length as the one layer thickness of $T_{\rm d}$-MoTe$_2$.} Figure~\ref{fig4}(a) displays the gate dependence of $T_c$ obtained from the 2 L sample. Here the gate voltage ($V_g$) is applied through a h-BN (34 nm in thickness) as a gate insulator. $T_c$ is successfully modulated by $V_g$, and at $V_g$ = 8 V it is larger by around 20 $\%$ compared with at $V_g$ = $-$8 V. In addition to the variation of $T_c$, the MCA signals are also modulated by $V_g$ [Fig.~\ref{fig4}(b)]. We find that not only the height of the peak but $B_{\rm peak}$ is also modulated by $V_g$, suggesting that the gating largely affects the vortex dynamics. Figure~\ref{fig4}(c) plots $\gamma$ as a function of $V_g$, showing the large variation of $\gamma$.

The gate voltage can modulate some parameters relevant to superconductivity, such as $T_c$, $B_{c2}$, the magnetic penetration length $\lambda$ and the coherence length $\xi$. \textcolor{black}{$\lambda$ is a characteristic scale for vortex-vortex interaction, whose crucial role was previously pointed out in the ratchet-like motion \cite{Nori, Palau}, whereas nonreciprocal signals in the superconducting state are largely affected by the variation of $T_c$ \cite{Wakatsuki, Hoshino, Itahashi2}. As the gating modulates these parameters in a complex manner, at present we cannot identify the dominant contribution to the large variation of $\gamma$. We hope that our results stimulate further theoretical as well as experimental investigations to reveal the role of the crystal symmetry for MCA and vortex dynamics in noncentrosymmetric superconductors.}

In conclusion, we showed giant superconducting nonreciprocal transport (MCA) in thin samples of the noncentrosymmetric superconductor $T_{\rm d}$-MoTe$_2$, which has only one mirror plane. We obtained 3.1 $\times$ 10$^6$ T$^{-1}$ A$^{-1}$ at 230 mK, one of the largest values of $\gamma$ recorded so far. The temperature dependence of $\gamma$ supports the ratchet-like motion of magnetic vortices as the origin of the nonreciprocal transport. The reduced symmetry of the crystal structure of $T_{\rm d}$-MoTe$_2$ may contribute to the large nonreciprocal signals. We also demonstrated gate modulation of the MCA in the superconducting state. In the 2 L $T_{\rm d}$-MoTe$_2$, we obtain a substantial modulation of $\gamma$ using a typical solid gate. Simultaneous demonstration of the gigantic MCA and its gate modulation in the superconducting state reveals that $T_{\rm d}$-MoTe$_2$ is a potential candidate for realizing electrically-tunable efficient superconducting rectification devices. 

We gratefully acknowledge M. Imai, S. Sasaki, H. Murofushi and S. Wang for their support in the experiments.
This project is financially supported in part by the JPSJ KAKENHI (Grant Number JP21H01022, JP21H04652, JP21K18181, JP21H05236, JP20H00354 and JP19H05790).

\appendix

\section{Thickness identification via atomic force microscope (AFM)}

\begin{figure}[tb!]
\begin{center}
\includegraphics[width=8cm,clip]{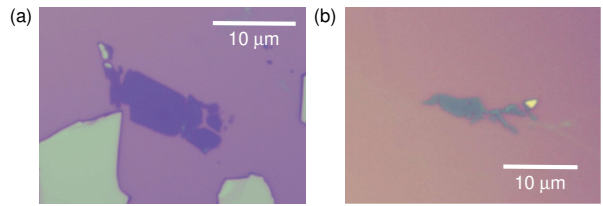}
\caption{Optical microscope images of thin $T_{\rm d}$-MoTe$_2$ flakes obtained via mechanical exfoliation.(a): 4 L (b): 2 L.}
\label{figMoTe2}
\end{center}
\end{figure}

\begin{figure}[b!]
\begin{center}
\includegraphics[width=8cm,clip]{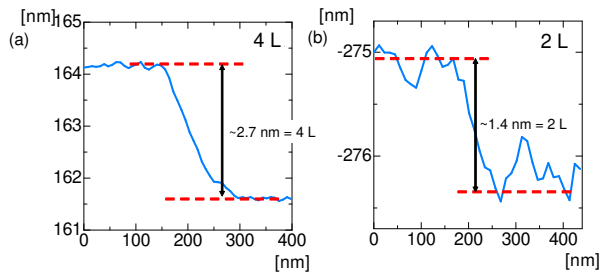}
\caption{Height profiles obtained from AFM scan of the 4 L (a) and 2 L (b) samples.}
\label{figAFM}
\end{center}
\end{figure}

Here we show the AFM data on the thickness of the thin $T_{\rm d}$-MoTe$_2$ flakes. Figures~\ref{figMoTe2} show some examples of exfoliated flakes a few layers in thickness on a SiO$_2$(285 nm)/Si substrate. As shown in Fig.~\ref{figAFM}(a) and (b), the two samples employed in this study are 4 L and 2 L. Note that AFM measurements are performed after the transport measurements, thus $T_{\rm d}$-MoTe$_2$ flakes are encapsulated by hBN.

\section{Determination of the crystal axes by Raman scattering}
\begin{figure*}[tb!]
\begin{center}
\includegraphics[width=14cm,clip]{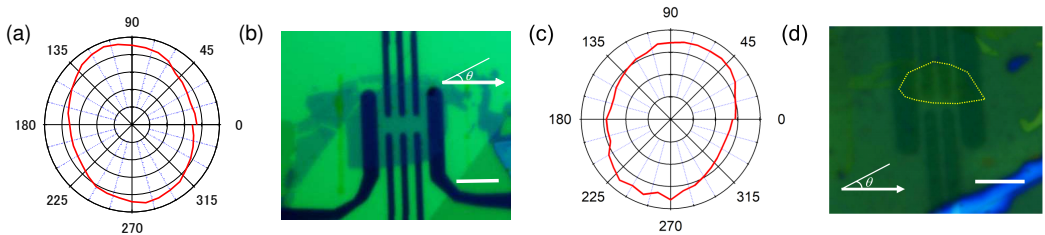}
\caption{(a) Angular dependence of the Raman intensity at the 163 cm$^{-1}$ Raman shift for the 4 L sample. (b) Optical image of the 4 L sample. The orientation of the sample corresponds to the orientation of the angle in (a). The scale bar is 5 $\mu$m. 
\textcolor{black}{(c) Angular dependence for the 2 L sample. (d) Sample picture of the 2 L sample, with 5 $\mu$m scale bar.}
}
\label{figRaman}
\end{center}
\end{figure*}

Since $T_{\rm d}$-MoTe$_2$ is a highly anisotropic material, it is important to determine the crystal axes of the sample and associate them with its transport properties. The polarization-angle dependence of the Raman intensity is a powerful tool for identifying the crystal axes \cite{Song2017, Zhou2017}. We perform Raman scattering measurements using HeNe laser light at 633 nm and measure the polarization-angle dependence of the Raman intensity at 163 cm$^{-1}$ Raman shift. Previous studies on angle-resolved Raman scattering measurements reported that the Raman intensity at 163 cm$^{-1}$ exhibits a maximum when the polarization is parallel to the Mo-zigzag chain ($b$-axis), and a minimum when it is perpendicular to it ($a$-axis). Figure~\ref{figRaman}(a) displays the angle dependence of the Raman intensity at 163 cm$^{-1}$ for the 4 L sample. Due to the reduced thickness of the sample in comparison with bulk, the otherwise typical two-lobe structure is deformed, but the in-plane anisotropy and the orientation of the axes are clearly discernible. The corresponding orientation of the 4 L sample is shown in Fig.~\ref{figRaman}(b). We can see that the electrodes are slightly misaligned from the direction of the principal crystal axes, which is the reason for a finite $R_{2 \omega}$ for $I \parallel a$ [$\equiv R_{2 \omega}(I \parallel a)$] shown in the main text. The misalignment angle is estimated to be around 15$^{\circ}$, leading to $\sin(-15^{\circ}) \sim -0.26$. This value is in close agreement with the ratio $R_{2 \omega}(I \parallel a)/R_{2 \omega}(I \parallel b) = -0.23$, taking into account that the peak (dip) is observed in the positive magnetic field for $R_{2 \omega}(I \parallel a)$ [$R_{2 \omega}(I \parallel b)]$.
\textcolor{black}{Figures~\ref{figRaman}(c) and (d) display the angular dependence of the Raman intensity and the sample picture for the 2 L sample. Misalignment between the electrodes and the crystal axes is larger, probably leading to a slight suppression of the nonreciprocal signals compared with those for the 4 L sample.}   

\begin{figure*}[htb!]
\begin{center}
\includegraphics[width=14cm,clip]{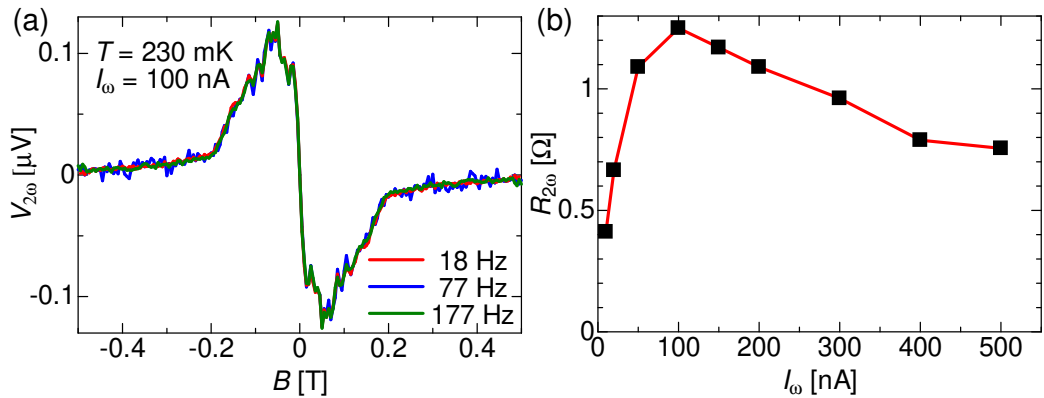}
\caption{(a) $R_{2 \omega}$ at different frequencies. No frequency dependencies are observed. (b) Driving current ($I_\omega$) dependence of $R_{2 \omega}$. $I_\omega$ = 100 nA provides the largest $R_{2 \omega}$. The data in (a) and (b) are both from the 4 L sample.}
\label{figfreq}
\end{center}
\end{figure*}



\section{Current and frequency dependence of the nonreciprocal resistance}
The main text shows $R_{2 \omega}$ as a function of $B$ with $I_\omega$ at 18 Hz. One may assume that the second harmonic signals are due to some spurious effect such as a capacitive coupling of the sample to the surrounding conductive environment. To rule out this possibility, we carry out $R_{2 \omega}$ measurements by driving $I_\omega$ at different frequencies. Figure~\ref{figfreq}(a) displays the results, demonstrating that $R_{2 \omega}$ signals are independent of the frequency of $I_\omega$. This is corroboration that $R_{2 \omega}$ signals we measure derive from the intrinsic transport properties of $T_{\rm d}$-MoTe$_2$, namely, magnetochiral anisotropy (MCA).

As mentioned in the main text, the efficiency of generating the nonreciprocal resistance is evaluated as $\gamma = 2 R_{2 \omega}/(R_\omega B I_\omega)$. Because of the definition of $\gamma$, one might naively think that $\gamma$ is divergent in the limit of $I_\omega \rightarrow 0$. However, this would be incorrect considering that the origin of the nonreciprocal transport is the ratchet-like motion of the magnetic vortices, because the vortices are not driven when the Lorenz force exerted by $I_\omega$ does not overcome the pinning force. On the other hand, superconductivity is suppressed if $I_\omega$ is too large. Therefore, it is expected that there is an intermediate value of $I_\omega$ which provides the largest signal of $R_{2 \omega}$. Indeed, as shown in Fig.~\ref{figfreq}(b), $R_{2 \omega}$ is suppressed as $I_\omega \rightarrow 0$, and there is an optimal value to obtain the largest $R_{2 \omega}$, which is 100 nA for the 4 L sample. $R_{2 \omega}$ diminishes when $I_\omega$ is larger than this value. Therefore, $I_\omega$ = 100 nA was used for the nonreciprocal transport measurements shown in the main text. Note that the optimum value of $I_\omega$ for the 2 L sample is 200 nA reflecting higher $T_c$.

\section{MCA for the 2 L ${\rm MoTe}_2$ sample and comparison of $\gamma$ as a function of $B$}
\begin{figure*}[tb!]
\begin{center}
\includegraphics[width=14cm,clip]{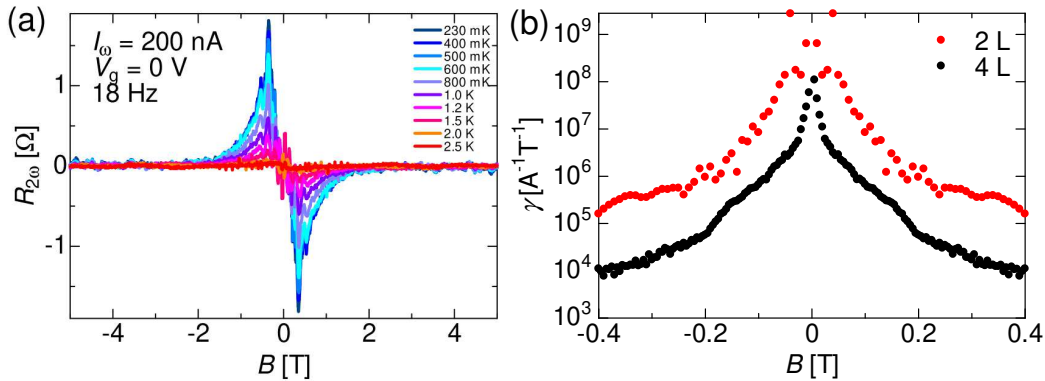}
\caption{(a) MCA signals obtained from the 2 L sample at different temperatures. The characteristics are similar to those for the 4 L sample. (b) Comparison of the relation between $\gamma$ and $B$ at 230 mK between the 2 L (at $V_g$ = 0) and 4 L sample.
}
\label{figbilayer}
\end{center}
\end{figure*}

Whereas the main text we principally shows the data from the 4 L sample, similar nonreciprocal transport results are also obtained for the 2 L sample. Figure~\ref{figbilayer}(a) shows $R_{2 \omega}$ curves taken at different temperatures from the 2 L sample. Increasing $R_{2 \omega}$ with decreasing temperature is visible explicitly, while the value of $B_{\rm peak}$ is different from that of the 4 L sample because of the higher $T_c$. 
\textcolor{black}{In the main text we discuss the value of $\gamma$ using the values of $R_{2 \omega}$ and $R_\omega$ at $B_{\rm peak}$. With this estimate, $\gamma$ for the 2 L sample is slightly smaller than the 4 L sample. By contrast, we can also plot $\gamma$ as a function of $B$ as we show in the bottom panel of Fig. 2(a). Figure~\ref{figbilayer}(b) displays the evolution of $\gamma$ with $B$ from the 2 L and 4 L samples. It is evident that at any $B$, $\gamma$ from the 2 L sample is larger than that from the 4 L sample. This indicates that the smaller $\gamma$ for the 2 L sample is mainly due to the larger $B_{\rm peak}$.}

\section{Vortex phase diagram and estimation of $U_0$}
\begin{figure*}[tb!]
\begin{center}
\includegraphics[width=14cm,clip]{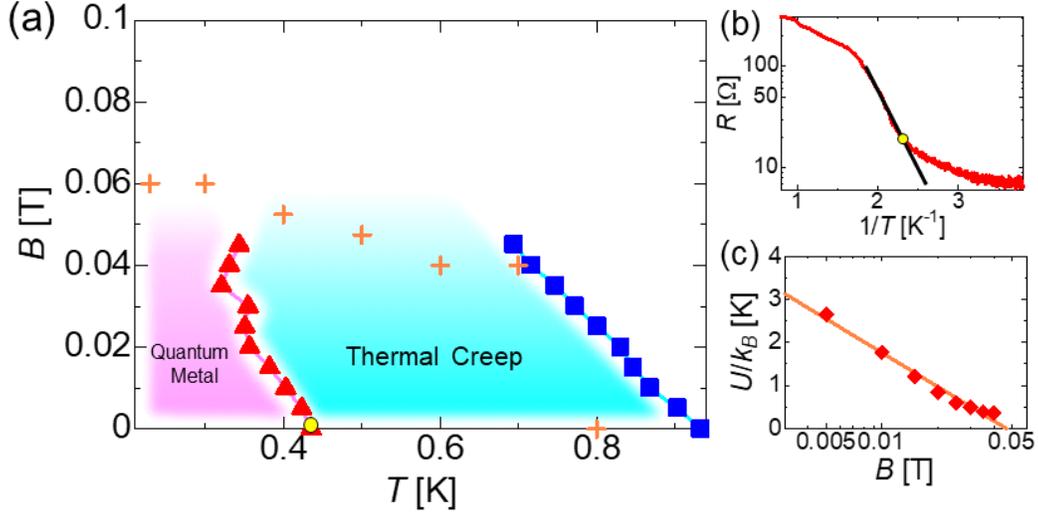}
\caption{(a) Vortex phase diagram obtained from the 4 L sample. Thermal creep (thermally activated flux flow) phase transits into the quantum metal phase at lower temperatures. Blue squares are determined from the upper critical field $B_{c2}$, and red triangles are obtained from the temperature dependence of the resistance under perpendicular magnetic field. Orange cross marks express the points at which $R_{2 \omega}$ takes a peak. All the cross marks apart from two of them at lower temperatures are inside the thermal creep phase, supporting the vortex dynamics plays a central role for giant nonreciprocal signals. (b) Arrhenius plot of the temperature dependence of the resistance. The boundary point between the quantum metal and thermal creep phase is determined as a point (the yellow point) at which the Arrhenius plot deviates from the exponential decay (black solid line). This yellow point corresponds to the yellow point in (a). (c) The slope shown by the solid line in (b) provides the activation energy $U(B)$ at different magnetic fields. From the slope of $U(B)/k_B$ as a function of $B$ (orange solid line), we can obtain $U_0$.} 
\label{vortex_phase}
\end{center}
\end{figure*}

Since giant MCA observed in $T_{\rm d}$-MoTe$_2$ is likely attributed to the ratchet-like motion of magnetic vortices, we show the vortex phase diagram in Fig.~\ref{vortex_phase}(a) to identify the vortex state at each temperature and magnetic field where we observe a large nonreciprocal signal. At lower temperatures and lower magnetic fields, vortices are in the "quantum metal" phase, in which a finite resistance remains even much below $T_c$ under a magnetic field \cite{Saito2015}. Vortices are in the thermal creep (or thermally assisted flux flow) regime at higher temperatures and higher magnetic fields up to $T_c$ and the upper critical magnetic field ($H_{c2}$), where vortices are mobile and can plastically flow under an excitation current. Ratchet-like motion of magnetic vortices is effective in the thermal creep regime. The experimental data points which separates the quantum metal phase and thermal creep phase are obtained following \cite{Saito2015}. Arrhenius plots of the temperature dependence of the resistance are prepared under a magnetic field, and the temperature at which the plot deviates from the exponential decay is defined as the boundary temperature [see Fig. \ref{vortex_phase}(b)]. Repeating this procedure for different magnetic fields provides sets of data ($T$,~$B$) for the boundary between the thermal creep and the quantum metal phase, as plotted in Fig.~\ref{vortex_phase}(a). The cross marks in Fig.~\ref{vortex_phase}(a) compose sets of temperature and magnetic field condition at which the peaks in $R_{2\omega}$ are observed. Most of the points are contained in the thermal creep region, except for the two points at lower temperature, giving another evidence that vortex dynamics affected by the ratchet-like potential plays a principle role for giant nonreciprocal signals. In the phase diagram it is visible that the quantum metal phase extends to relatively larger magnetic fields at lower temperatures. The remaining two points overlap this region, indicating that quantum effects may affect the nonreciprocal signals. This explains the suppression of $\gamma$ obtained from experiments in comparison to the theoretical fit based on the ratchet-like motion of magnetic vortices as we discuss in the main text.

\textcolor{black}{Temperature ($T$) dependence of the resistance under a magnetic field also enables to estimate the potential height $U_0$. In the thermal creep regime, the resistance $R$ depends on the activation energy $U(B)$ as $R = R_0 \exp[-U(B)/k_B T]$ with the normal state resistance $R_0$ \cite{Saito2015}. Thus the slope in a logarithmic plot of $R$ with $1/T$ provides $U(B)$ [see also Fig.~\ref{vortex_phase}(b)]. Measuring $T$-dependence of $R$ under different $B$ provides the relation between $U(B)$ and $B$, from which we can obtain $U_0$ = 0.10(0.56) meV by using the relation $U(B) = U_0 \ln(B_0/B)$ for the 4 L (2 L) samples [Fig.~\ref{vortex_phase}(c)] \cite{Saito2015}.}

\textcolor{black}{\section{Axes dependence of the nonreciprocal signal}}
\textcolor{black}{In the main text we show the experimental results of MCA from the 2 L and 4 L samples. While these samples provide a sufficient amount of data sets, the crystal axes are not perfectly aligned parallel to the current direction, making it difficult to identify the axes dependence of the nonreciprocal signals. Here, we provide additional data from other samples whose crystal axes are almost parallel to the current direction. Figures~\ref{axes_dep}(a)-(c) are from another 4 L sample (4 L$\#$2) and the current direction is parallel to the $b$-axis ($I \parallel b$). The crystal axes are determined from the Raman intensity profile shown in Fig.~\ref{axes_dep}(b). As seen in Fig.~\ref{axes_dep}(c), clear peak and dip structures are observed. Figures~\ref{axes_dep}(d)-(f) display the second harmonic signal from the 6 L sample where $I \parallel a$. We can easily find in Fig.~\ref{axes_dep}(f) that while slowly oscillating background is visible, no peak and dip structures typical for MCA are observed. This also corroborates that peak and dip structures that we observe in the other samples arise from MCA.}

\section{Other possible effects to generate nonreciprocal signals}
While we have shown a number of additional experimental data which support MCA as the origin of the giant nonreciprocal signals, some may still wonder other effects can be considered to explain those nonreciprocal resistances. Let us rule out some other possibilities as an origin of the second harmonic signals.

\subsection{Thermal effects}
If there existed a thermal gradient $\bm{\nabla} T$, the Nernst effect would generate an electric field $\mathbf{E} \propto \mathbf{B} \times \bm{\nabla} T$ under a magnetic field $B$. The thermal gradient may be due to the Joule heating effect, therefore $\bm{\nabla} T \propto \mathbf{j}^2$, where $\mathbf{j}$ is current density, and $\bm{\nabla} T$ should be parallel to $\mathbf{j}$. Indeed, such a superconducting Nernst effect which generates a second harmonic voltage ($V_{2 \omega}$) was observed in another van der Waals superconductor NbSe$_2$, using a thermal gradient driven by a heater mounted close to the sample \cite{Li2020}. 
$V_{2 \omega}$ exhibits similar magnetic field dependence as those observed in our samples, derived from the (vortex) Nernst effect \cite{Behnia2016}. We can rule out the Nernst effect as a possible origin of MCA due to the following reasons: i) Temperature gradient assumed above should not exist considering that our $T_{\rm d}$-MoTe$_2$ is highly crystalline so that excitation currents pass almost homogeneously through the sample. ii) $V_{2 \omega}$ from the Nernst effect is reduced to zero at lower temperatures, opposite to our observations. Furthermore, the Nernst signal persists even above $T_c$, in contrast to the suppressed nonreciprocal signals above $T_c$. iii) $V_{2 \omega}$ induced by the Joule heating should be orthogonal to $\mathbf{j}$, inconsistent with our observations of large longitudinal nonreciprocal signals. 
\begin{figure*}[tb!]
\begin{center}
\includegraphics[width=14cm,clip]{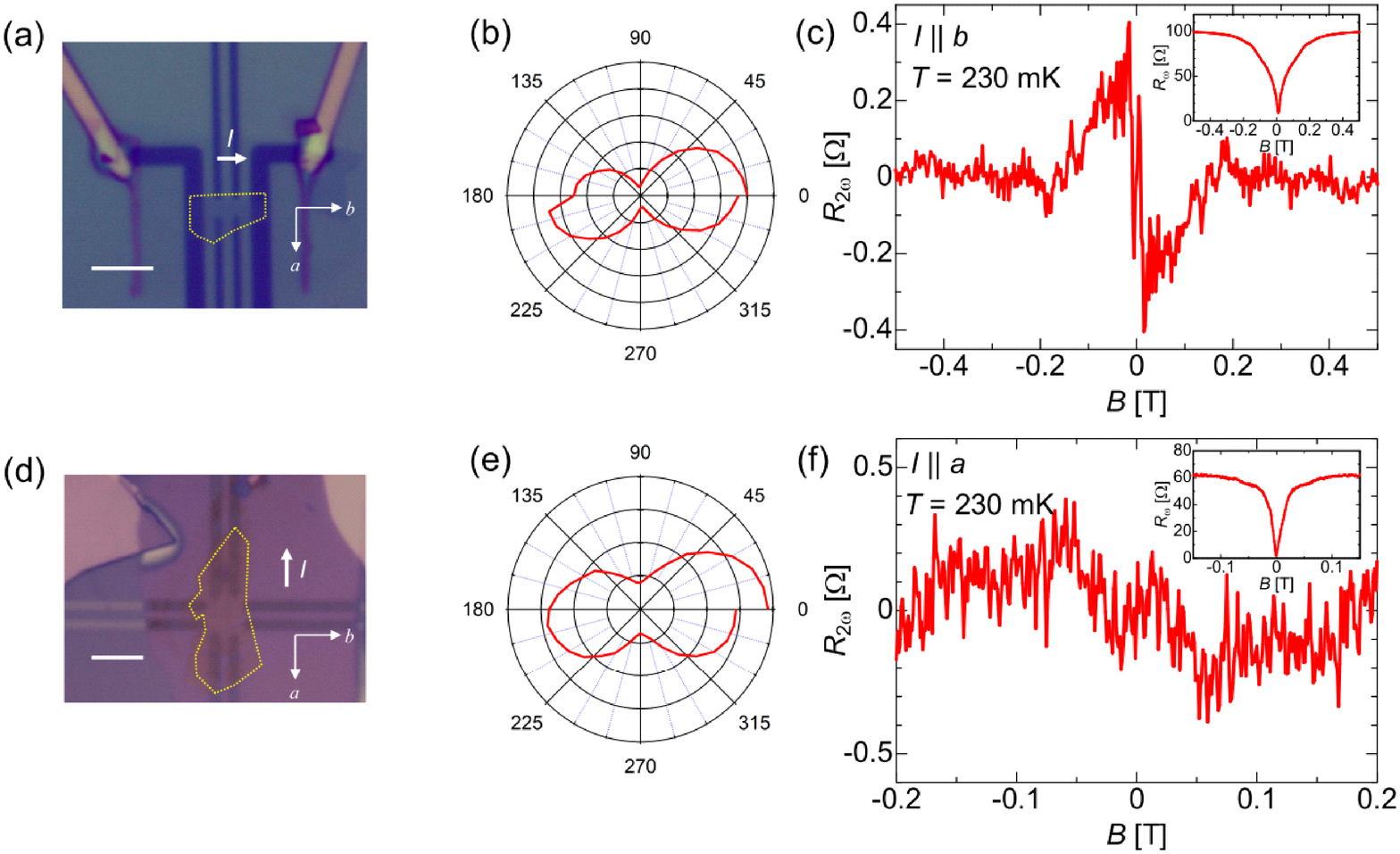}
\caption{\textcolor{black}{(a) Optical microscope image of the 4 L$\#$2 device. The directions of the crystal axes and excitation current are denoted. (b) Angular dependence of the Raman scattering intensity. The sample is set as shown in (a), thus 0 degree corresponds to the $b$-axis of the sample. (c) $R_{2 \omega}$ signal as a function of perpendicular magnetic field $B$ for $I \parallel b$. Large peak and dip structures are clearly visible. Inset shows the liner resistance $R_\omega$ simultaneously measured with $R_{2 \omega}$. (d) Optical microscope image of another sample with 6 L in thickness. Here the direction of a current $I$ is parallel to the $a$-axis, determined from the angular dependence of the Raman scattering intensity shown in (e). (f) $R_{2 \omega}$ for $I \parallel a$. There is a slowly oscillating background, but no peak and dip are observed. $R_\omega$ signal is displayed in the inset. Note that because of the problems in the electrodes, we cannot perform similar measurements driving $I$ parallel to the $b$-axis in this sample. In (a) and (d) $T_{\rm d}$-MoTe$_2$ is highlighted by a yellow dotted line, and the scale bar corresponds to 5 $\mu$m.}}
\label{axes_dep}
\end{center}
\end{figure*} 

\subsection{Geometrical effects}
Vortex rectification effect induced by the asymmetric sample geometry (e.g. asymmetric edge shape of the sample) was previously proposed \cite{Vodolazov2005} and experimentally confirmed \cite{Cerbu2013, Ji2016}. The main idea is that since magnetic vortices always enter from the edge (or surface in the case of three dimensional superconductors) of the sample and cannot nucleate inside the superconductor, the edge asymmetry between the opposite sides of the sample generates the asymmetric surface potential barrier for vortices. Because the edge selected for the vortex entry depends on the polarity of the current, the inequivalency in the potential barrier between the edges leads to the different current condition for the vortex entry thus rectification. One may wonder that the nonreciprocal signals observed in this study are attributed to the asymmetry between the edges of the sample. It is true that slight asymmetry between the edges is inevitable in our samples employing mechanically exfoliated $T_{\rm d}$-MoTe$_2$ flakes. If the edge asymmetry is a dominant contribution in our samples, larger nonreciprocal signals are expected for samples with more asymmetric edges. We do not see the correlation between the edge asymmetry and the amplitude of the nonreciprocal signal for different samples with different edge shapes. As an example, the edge asymmetry is more peculiar in the sample shown in Fig.~\ref{axes_dep}(d) than (a), but the nonreciprocal signal is much suppressed. Moreover, in our samples electrodes are always aligned symmetrically, ruling out the possibility of the asymmetry induced by electrodes. Therefore, we can draw a conclusion that vortex rectification due to the geometrical asymmetry in the sample is not an origin for the giant nonreciprocal signals.

\textcolor{black}{
\subsection{Surface barrier effect}}

\textcolor{black}{In relation to the geometrical effect, one may argue that the asymmetry in the surface barrier for magnetic vortices is the origin of MCA. As for the dynamics of magnetic vortices in superconductors, we must consider two effects: Bulk pinning and surface barrier. The important point is which effect is dominant in a certain condition. We highlight that the surface barrier effect plays a major role in the vortex dynamics only when the temperature is close to $T_c$, at which the bulk pinning becomes extremely weak. This important point has been already demonstrated by many previous studies \cite{SB1, SB3, SB4, Zeldov}. This behavior is inconsistent with our observation, indicating that bulk pinning effect is dominant in our samples.}




\section{Theoretical description using ratchet model}

\begin{figure}[b!]
\begin{center}
\includegraphics[width=8cm,clip]{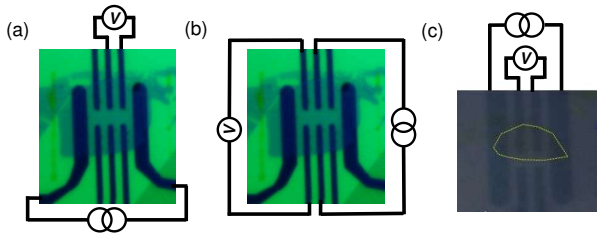}
\caption{(a) The current and voltage probes configuration for $I \parallel a$ in the 4 L sample. (b) The current and voltage probes configuration for $I \parallel b$ in the 4 L sample. (c) The current and voltage probes configuration in the 2 L sample. }
\label{current_dist}
\end{center}
\end{figure} 

\begin{figure*}[tb!]
\begin{center}
\includegraphics[width=14cm,clip]{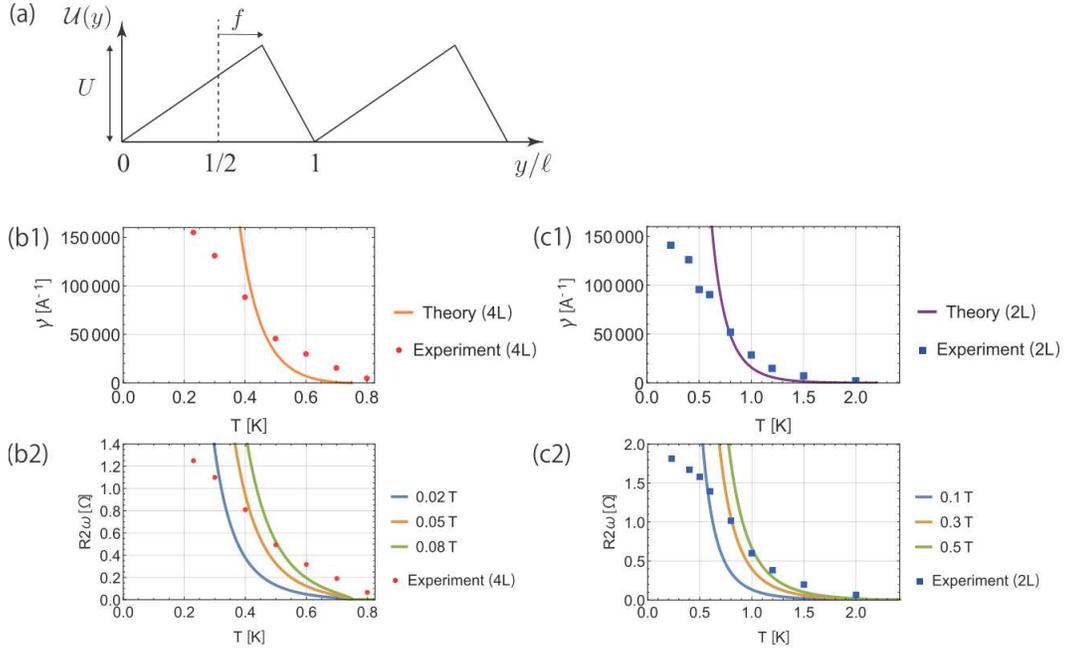}
\caption{
(a) Spatial dependence of the ratchet potential, where $f$ ($0\leq f\leq 1/2$) controls the asymmetry.
(b) Temperature dependence of (b1) $\gm'=\gm B$, and (b2) $R_{2\omega}$ for the 4 L sample.
Analogous plots for the 2 L sample are shown in (c1,c2).
}
\label{figTheor}
\end{center}
\end{figure*}

Here we describe the nonreciprocal transport signal originating from the vortex motion \cite{Hoshino, Itahashi2} .
For simplicity, we regard the vortex as a point particle.
The relation between velocity and force is written as
\begin{align}
v_{x} &= q_{1xx} F_x + q_{2xxy} F_x F_y 
\\
v_{y} &= q_{1yy} F_y + q_{2yxx} F_x^2 + q_{2yyy} F_y^2
\end{align}
where the mirror symmetry along the $y$ axis is assumed (for a trigonal symmetry, there are the relations $q_{1xx}=q_{1yy}$ and $q_{2xxy}=-2q_{2yxx}=2q_{2yyy}$).
If we take the configuration with $F_x=0$, only the $y$ direction is involved.
Hence we consider the one-dimensional equation of motion for the estimation of $q_{2yyy}$:
\begin{align}
    \eta \dot y = F_y -\frac{\partial \mathcal U(y)}{\partial y} + \xi(t)
\end{align}
where $ \displaystyle \eta = \frac{1.45 \pi \hbar^2\sigma_{\rm n}}{2e^2\xi^2}$ is a damping coefficient 
with the normal conductivity $\sigma_{\rm n}$ and in-plane coherence length $\xi$ \cite{Kopnin}.
The random force $\xi (t)$ represents a thermal noise satisfying $\la \xi (t) \xi(t') \ra = 2\eta k_{\rm B}T\delta(t-t')$ where the bracket indicates the random average.
The asymmetric potential $\mathcal U(y)$ for vortex is responsible for the non-reciprocal transport signal.
We take the periodic potential of the height $U$ and the periodicity $\ell$ given in Fig.~\ref{figTheor}(a).
The response coefficients are given by $\displaystyle q_{1yy}=\frac{1}{\eta} g_1(\beta U)$ and $\displaystyle q_{2yyy} = \frac{\beta \ell}{\eta} g_2(\beta U)$.
The functional forms of $g_{1,2}$ are explicitly given by
\begin{align}
&g_1(x) = \frac{x^2}{2(\cosh x-1)},\\
&g_2(x) = \frac{f \, x\, (4+x^2-4\cosh x+x\sinh x)}{2(\cosh x-1)^2}.
\end{align}
The parameter $f$ ($\leq \tfrac 1 2$) controls the asymmetry of the potential.

The force acting on the vortex is given by $F_y=j \phi_0^*$ where $j$ is a current density along $x$ direction and $\displaystyle \phi_0^* = \frac{h}{2|e|}$ is the flux quantum for superconductors.
The number density of vortices is given by $\displaystyle n=\frac{B}{\phi_0^*}$.
The voltage along $x$ direction is then given by the Josephson relation
$V_x = \phi_0^* L n v_y = R_1I + R_2I$ where $I=jW$ is a current with the sample width $W$ and length $L$.
The linear and non-linear transport coefficients are 
\begin{align}
R_1 &= \frac{\phi_0^* L B}{\eta W} g_1(\beta U)
,\ \ \ 
R_2 = \frac{(\phi_0^*)^2 ILB\ell}{\eta k_{\rm B}TW^2} g_2(\beta U)
.
\end{align}
Since both the signals are proportional to the vortex number density, their ratio is written in a simple form 
\begin{align}
\gm' &= \frac{R_{2}}{R_{1} I} 
= \frac{\phi_0^* \ell}{Wk_{\rm B}T} 
\cdot \frac{g_2(\beta U)}{g_1(\beta U)}
\end{align}
which is determined from the profile of the potential of vortices.
Note also the relations $R_{\omega}=R_1$ and $R_{2\omega}=R_2/2$.

Now we discuss characteristic parameters used in this model.
For the characteristic length $\ell$, we consider the value of the magnetic field $B_{\rm pin}$ at which all the pinning centers are occupied at low temperature.
The length scale is then given by $\ell(T=0) \sim \sqrt{\phi_0^*/B_{\rm pin}}$.
The potential height $U$ is estimated by the critical current density $j_{\rm c}$ at small magnetic field by the relation $U=j_c \phi_0^*\ell (\tfrac 1 2 + f)$.
We also consider the temperature dependence of these parameters since the size of vortices changes as the coherence length varies with increasing temperature.
We assume the temperature dependence of $\displaystyle U(T)\sim U_0 \Big( \frac{T_c-T}{T_c} \Big)^{\al}$ and $\displaystyle 
\ell(T) \sim \ell_0 \Big( \frac{T_c-T}{T_c} \Big)^{\al-\frac 1 2}$, which results in $j_c \propto \sqrt{T_c-T}$ ($\alpha=1$ is used in \cite{Hamamoto}).
Since the microscopic origin of the vortex potential is not clear at present, 
here we take $\alpha$ as a parameter phenomenologically for a better fit to the experimental data.
The temperature dependence of $U$ is more strongly reflected in the signal compared to that of $\ell$, because $\beta U$ is the argument of the non-linear function $g_{2}$.
 
Now we estimate the magnitude of the non-linear transport signal.
Since the nonreciprocal signals in magnetic field dependence is maximized in a moving vortex regime, we assume the expression for a small pinning potential
$g_1(\beta U) \sim 1$ and $\displaystyle g_2(\beta U) \sim f \frac{ (\beta U)^3}{180}$, with which the vortices are not fixed in the pinning potential.
We use the parameters for the 4 L (2 L) sample such as
$W=5\ (2.5)\ \mu{\rm m}$, $T_c=0.75\ (2.2)$ K, $B_{\rm pin}(T=0) \sim 0.025\ (0.2)$ T, and $I_c(T=0)=j_cW\sim 200\ (500)$ nA. 
$B_{\rm pin}$ is estimated from the magnetic-field dependence of the resistance, and $I_c$ from the current-voltage characteristics under a magnetic field measured for each sample. 
In order to be compatible with experiments, we choose the vortex potential parameters as $f=0.15$ and $\alpha = 0.5$, which are used for both the 4 L and 2 L samples.
With these parameters, the pinning potential height at zero temperature is estimated as $U_0 \sim 0.1\ (0.17)$ meV. Note that the estimated values of $U_0$ are close to those experimentally obtained from the temperature dependence of the resistance under a perpendicular magnetic field \textcolor{black}{(see Appendix E)}.  
Since $\gm'$ is proportional to $(T_c-T)^{4\al-\frac 1 2}$ in our model, the temperature dependence becomes more convex downward if we take larger $\al$.

The temperature dependence of $\gm' = \gamma B$ is shown in Fig.~\ref{figTheor} (b1) for the 4 L sample and in (c1) for the 2 L sample.
The vortex ratchet model with two adjustable parameters $\alpha$ and $f$ reproduces the magnitude of signals observed in experiments.
We note that the present model can be justified in the middle temperature range.
For high temperature range $T\gtrsim T_{c}$, the vortex picture should be replaced by superconducting fluctuation mechanism and/or normal contribution.
At low temperatures, on the other hand, a quantum effect on ratchet-like motion \cite{Hamamoto} is needed for more accurate estimate, as supported by the vortex phase diagram discussed above.


With the above setup, we also estimate the values of $R_{2\omega}$ by using
the normal resistance $R_{\rm n} = \frac{L}{W}\sg_{\rm n} \sim 330\ (380)\ \Omega$, the coherence length $\xi(T) \sim \xi_0 \sqrt{T_c/(T_c-T)}$ with $\xi_0\sim 45\ (24)$ nm, and the current $I=100\ (200)$ nA.
The results are shown in Fig.~\ref{figTheor}(b2) for the 4 L sample and (c2) for the 2 L sample.
Since the experimental data plotted here are obtained using different magnetic fields ($0.04$--$0.05$ T for 4 L and $0.25$--$0.42$ T for 2 L), we show several lines for different magnetic fields.
We confirm that the present model can roughly reproduce the magnitude $R_{2\omega}$ in a middle temperature range below $T_c$.
$R_{2\omega}$ is decreasing function of temperature, which
is also consistent with experiments.

\bibliography{prr}

\providecommand{\noopsort}[1]{}\providecommand{\singleletter}[1]{#1}%
\begin{thebibliography}{48}%
\makeatletter
\providecommand \@ifxundefined [1]{%
 \@ifx{#1\undefined}
}%
\providecommand \@ifnum [1]{%
 \ifnum #1\expandafter \@firstoftwo
 \else \expandafter \@secondoftwo
 \fi
}%
\providecommand \@ifx [1]{%
 \ifx #1\expandafter \@firstoftwo
 \else \expandafter \@secondoftwo
 \fi
}%
\providecommand \natexlab [1]{#1}%
\providecommand \enquote  [1]{``#1''}%
\providecommand \bibnamefont  [1]{#1}%
\providecommand \bibfnamefont [1]{#1}%
\providecommand \citenamefont [1]{#1}%
\providecommand \href@noop [0]{\@secondoftwo}%
\providecommand \href [0]{\begingroup \@sanitize@url \@href}%
\providecommand \@href[1]{\@@startlink{#1}\@@href}%
\providecommand \@@href[1]{\endgroup#1\@@endlink}%
\providecommand \@sanitize@url [0]{\catcode `\\12\catcode `\$12\catcode
  `\&12\catcode `\#12\catcode `\^12\catcode `\_12\catcode `\%12\relax}%
\providecommand \@@startlink[1]{}%
\providecommand \@@endlink[0]{}%
\providecommand \url  [0]{\begingroup\@sanitize@url \@url }%
\providecommand \@url [1]{\endgroup\@href {#1}{\urlprefix }}%
\providecommand \urlprefix  [0]{URL }%
\providecommand \Eprint [0]{\href }%
\providecommand \doibase [0]{https://doi.org/}%
\providecommand \selectlanguage [0]{\@gobble}%
\providecommand \bibinfo  [0]{\@secondoftwo}%
\providecommand \bibfield  [0]{\@secondoftwo}%
\providecommand \translation [1]{[#1]}%
\providecommand \BibitemOpen [0]{}%
\providecommand \bibitemStop [0]{}%
\providecommand \bibitemNoStop [0]{.\EOS\space}%
\providecommand \EOS [0]{\spacefactor3000\relax}%
\providecommand \BibitemShut  [1]{\csname bibitem#1\endcsname}%
\let\auto@bib@innerbib\@empty
\bibitem [{\citenamefont {Ando}\ \emph {et~al.}(2020)\citenamefont {Ando},
  \citenamefont {Miyasaka}, \citenamefont {Li}, \citenamefont {Ishizuka},
  \citenamefont {Arakawa}, \citenamefont {Shiota}, \citenamefont {Moriyama},
  \citenamefont {Yanase},\ and\ \citenamefont {Ono}}]{SCDiode_Ono}%
  \BibitemOpen
  \bibfield  {author} {\bibinfo {author} {\bibfnamefont {F.}~\bibnamefont
  {Ando}}, \bibinfo {author} {\bibfnamefont {Y.}~\bibnamefont {Miyasaka}},
  \bibinfo {author} {\bibfnamefont {T.}~\bibnamefont {Li}}, \bibinfo {author}
  {\bibfnamefont {J.}~\bibnamefont {Ishizuka}}, \bibinfo {author}
  {\bibfnamefont {T.}~\bibnamefont {Arakawa}}, \bibinfo {author} {\bibfnamefont
  {Y.}~\bibnamefont {Shiota}}, \bibinfo {author} {\bibfnamefont
  {T.}~\bibnamefont {Moriyama}}, \bibinfo {author} {\bibfnamefont
  {Y.}~\bibnamefont {Yanase}},\ and\ \bibinfo {author} {\bibfnamefont
  {T.}~\bibnamefont {Ono}},\ }\bibfield  {title} {\bibinfo {title} {Observation
  of superconducting diode effect},\ }\href
  {https://doi.org/10.1038/s41586-020-2590-4} {\bibfield  {journal} {\bibinfo
  {journal} {Nature}\ }\textbf {\bibinfo {volume} {584}},\ \bibinfo {pages}
  {373} (\bibinfo {year} {2020})}\BibitemShut {NoStop}%
\bibitem [{\citenamefont {Wu}\ \emph {et~al.}(2022)\citenamefont {Wu},
  \citenamefont {Wang}, \citenamefont {Xu}, \citenamefont {Sivakumar},
  \citenamefont {Pasco}, \citenamefont {Filippozzi}, \citenamefont {Parkin},
  \citenamefont {Zeng}, \citenamefont {McQueen},\ and\ \citenamefont
  {Ali}}]{SCDiode_Ali}%
  \BibitemOpen
  \bibfield  {author} {\bibinfo {author} {\bibfnamefont {H.}~\bibnamefont
  {Wu}}, \bibinfo {author} {\bibfnamefont {Y.}~\bibnamefont {Wang}}, \bibinfo
  {author} {\bibfnamefont {Y.}~\bibnamefont {Xu}}, \bibinfo {author}
  {\bibfnamefont {P.~K.}\ \bibnamefont {Sivakumar}}, \bibinfo {author}
  {\bibfnamefont {C.}~\bibnamefont {Pasco}}, \bibinfo {author} {\bibfnamefont
  {U.}~\bibnamefont {Filippozzi}}, \bibinfo {author} {\bibfnamefont {S.~S.~P.}\
  \bibnamefont {Parkin}}, \bibinfo {author} {\bibfnamefont {Y.-J.}\
  \bibnamefont {Zeng}}, \bibinfo {author} {\bibfnamefont {T.}~\bibnamefont
  {McQueen}},\ and\ \bibinfo {author} {\bibfnamefont {M.~N.}\ \bibnamefont
  {Ali}},\ }\bibfield  {title} {\bibinfo {title} {The field-free josephson
  diode in a van der waals heterostructure},\ }\href
  {https://doi.org/10.1038/s41586-022-04504-8} {\bibfield  {journal} {\bibinfo
  {journal} {Nature}\ }\textbf {\bibinfo {volume} {604}},\ \bibinfo {pages}
  {653} (\bibinfo {year} {2022})}\BibitemShut {NoStop}%
\bibitem [{\citenamefont {Bauriedl}\ \emph {et~al.}(2022)\citenamefont
  {Bauriedl}, \citenamefont {B{\"a}uml}, \citenamefont {Fuchs}, \citenamefont
  {Baumgartner}, \citenamefont {Paulik}, \citenamefont {Bauer}, \citenamefont
  {Lin}, \citenamefont {Lupton}, \citenamefont {Taniguchi}, \citenamefont
  {Watanabe}, \citenamefont {Strunk},\ and\ \citenamefont
  {Paradiso}}]{SCDiode_NbSe2}%
  \BibitemOpen
  \bibfield  {author} {\bibinfo {author} {\bibfnamefont {L.}~\bibnamefont
  {Bauriedl}}, \bibinfo {author} {\bibfnamefont {C.}~\bibnamefont {B{\"a}uml}},
  \bibinfo {author} {\bibfnamefont {L.}~\bibnamefont {Fuchs}}, \bibinfo
  {author} {\bibfnamefont {C.}~\bibnamefont {Baumgartner}}, \bibinfo {author}
  {\bibfnamefont {N.}~\bibnamefont {Paulik}}, \bibinfo {author} {\bibfnamefont
  {J.~M.}\ \bibnamefont {Bauer}}, \bibinfo {author} {\bibfnamefont {K.-Q.}\
  \bibnamefont {Lin}}, \bibinfo {author} {\bibfnamefont {J.~M.}\ \bibnamefont
  {Lupton}}, \bibinfo {author} {\bibfnamefont {T.}~\bibnamefont {Taniguchi}},
  \bibinfo {author} {\bibfnamefont {K.}~\bibnamefont {Watanabe}}, \bibinfo
  {author} {\bibfnamefont {C.}~\bibnamefont {Strunk}},\ and\ \bibinfo {author}
  {\bibfnamefont {N.}~\bibnamefont {Paradiso}},\ }\bibfield  {title} {\bibinfo
  {title} {Supercurrent diode effect and magnetochiral anisotropy in few-layer
  $\mathrm{NbSe}_2$},\ }\href {https://doi.org/10.1038/s41467-022-31954-5}
  {\bibfield  {journal} {\bibinfo  {journal} {Nature Communications}\ }\textbf
  {\bibinfo {volume} {13}},\ \bibinfo {pages} {4266} (\bibinfo {year}
  {2022})}\BibitemShut {NoStop}%
\bibitem [{\citenamefont {Tokura}\ and\ \citenamefont
  {Nagaosa}(2018)}]{Tokura}%
  \BibitemOpen
  \bibfield  {author} {\bibinfo {author} {\bibfnamefont {Y.}~\bibnamefont
  {Tokura}}\ and\ \bibinfo {author} {\bibfnamefont {N.}~\bibnamefont
  {Nagaosa}},\ }\bibfield  {title} {\bibinfo {title} {Nonreciprocal responses
  from non-centrosymmetric quantum materials},\ }\href
  {https://doi.org/10.1038/s41467-018-05759-4} {\bibfield  {journal} {\bibinfo
  {journal} {Nature Communications}\ }\textbf {\bibinfo {volume} {9}},\
  \bibinfo {pages} {3740} (\bibinfo {year} {2018})}\BibitemShut {NoStop}%
\bibitem [{\citenamefont {Ideue}\ \emph {et~al.}(2017)\citenamefont {Ideue},
  \citenamefont {Hamamoto}, \citenamefont {Koshikawa}, \citenamefont {Ezawa},
  \citenamefont {Shimizu}, \citenamefont {Kaneko}, \citenamefont {Tokura},
  \citenamefont {Nagaosa},\ and\ \citenamefont {Iwasa}}]{Ideue}%
  \BibitemOpen
  \bibfield  {author} {\bibinfo {author} {\bibfnamefont {T.}~\bibnamefont
  {Ideue}}, \bibinfo {author} {\bibfnamefont {K.}~\bibnamefont {Hamamoto}},
  \bibinfo {author} {\bibfnamefont {S.}~\bibnamefont {Koshikawa}}, \bibinfo
  {author} {\bibfnamefont {M.}~\bibnamefont {Ezawa}}, \bibinfo {author}
  {\bibfnamefont {S.}~\bibnamefont {Shimizu}}, \bibinfo {author} {\bibfnamefont
  {Y.}~\bibnamefont {Kaneko}}, \bibinfo {author} {\bibfnamefont
  {Y.}~\bibnamefont {Tokura}}, \bibinfo {author} {\bibfnamefont
  {N.}~\bibnamefont {Nagaosa}},\ and\ \bibinfo {author} {\bibfnamefont
  {Y.}~\bibnamefont {Iwasa}},\ }\bibfield  {title} {\bibinfo {title} {Bulk
  rectification effect in a polar semiconductor},\ }\href
  {https://doi.org/10.1038/nphys4056} {\bibfield  {journal} {\bibinfo
  {journal} {Nature Physics}\ }\textbf {\bibinfo {volume} {13}},\ \bibinfo
  {pages} {578} (\bibinfo {year} {2017})}\BibitemShut {NoStop}%
\bibitem [{\citenamefont {Legg}\ \emph {et~al.}(2022)\citenamefont {Legg},
  \citenamefont {Rossler}, \citenamefont {Munning}, \citenamefont {Fan},
  \citenamefont {Breunig}, \citenamefont {Bliesener}, \citenamefont {Lippertz},
  \citenamefont {Uday}, \citenamefont {Taskin}, \citenamefont {Loss},
  \citenamefont {Klinovaja},\ and\ \citenamefont {Ando}}]{Ando}%
  \BibitemOpen
  \bibfield  {author} {\bibinfo {author} {\bibfnamefont {H.~F.}\ \bibnamefont
  {Legg}}, \bibinfo {author} {\bibfnamefont {M.}~\bibnamefont {Rossler}},
  \bibinfo {author} {\bibfnamefont {F.}~\bibnamefont {Munning}}, \bibinfo
  {author} {\bibfnamefont {D.}~\bibnamefont {Fan}}, \bibinfo {author}
  {\bibfnamefont {O.}~\bibnamefont {Breunig}}, \bibinfo {author} {\bibfnamefont
  {A.}~\bibnamefont {Bliesener}}, \bibinfo {author} {\bibfnamefont
  {G.}~\bibnamefont {Lippertz}}, \bibinfo {author} {\bibfnamefont
  {A.}~\bibnamefont {Uday}}, \bibinfo {author} {\bibfnamefont {A.~A.}\
  \bibnamefont {Taskin}}, \bibinfo {author} {\bibfnamefont {D.}~\bibnamefont
  {Loss}}, \bibinfo {author} {\bibfnamefont {J.}~\bibnamefont {Klinovaja}},\
  and\ \bibinfo {author} {\bibfnamefont {Y.}~\bibnamefont {Ando}},\ }\bibfield
  {title} {\bibinfo {title} {Giant magnetochiral anisotropy from
  quantum-confined surface states of topological insulator nanowires},\ }\href
  {https://doi.org/10.1038/s41565-022-01124-1} {\bibfield  {journal} {\bibinfo
  {journal} {Nature Nanotechnology}\ }\textbf {\bibinfo {volume} {17}},\
  \bibinfo {pages} {696} (\bibinfo {year} {2022})}\BibitemShut {NoStop}%
\bibitem [{\citenamefont {Villegas}\ \emph
  {et~al.}(2003{\natexlab{a}})\citenamefont {Villegas}, \citenamefont
  {Savel'ev}, \citenamefont {Nori}, \citenamefont {Gonzalez}, \citenamefont
  {Anguita}, \citenamefont {Garc{\'i}a},\ and\ \citenamefont
  {Vicent}}]{Ratchet1}%
  \BibitemOpen
  \bibfield  {author} {\bibinfo {author} {\bibfnamefont {J.~E.}\ \bibnamefont
  {Villegas}}, \bibinfo {author} {\bibfnamefont {S.}~\bibnamefont {Savel'ev}},
  \bibinfo {author} {\bibfnamefont {F.}~\bibnamefont {Nori}}, \bibinfo {author}
  {\bibfnamefont {E.~M.}\ \bibnamefont {Gonzalez}}, \bibinfo {author}
  {\bibfnamefont {J.~V.}\ \bibnamefont {Anguita}}, \bibinfo {author}
  {\bibfnamefont {R.}~\bibnamefont {Garc{\'i}a}},\ and\ \bibinfo {author}
  {\bibfnamefont {J.~L.}\ \bibnamefont {Vicent}},\ }\bibfield  {title}
  {\bibinfo {title} {A superconducting reversible rectifier that controls the
  motion of magnetic flux quanta},\ }\href
  {https://doi.org/10.1126/science.1090390} {\bibfield  {journal} {\bibinfo
  {journal} {Science}\ }\textbf {\bibinfo {volume} {302}},\ \bibinfo {pages}
  {1188} (\bibinfo {year} {2003}{\natexlab{a}})}\BibitemShut {NoStop}%
\bibitem [{\citenamefont {Lee}\ \emph {et~al.}(1999)\citenamefont {Lee},
  \citenamefont {Jank{\'o}}, \citenamefont {Der{\'e}nyi},\ and\ \citenamefont
  {Barab{\'a}si}}]{Ratchet2}%
  \BibitemOpen
  \bibfield  {author} {\bibinfo {author} {\bibfnamefont {C.-S.}\ \bibnamefont
  {Lee}}, \bibinfo {author} {\bibfnamefont {B.}~\bibnamefont {Jank{\'o}}},
  \bibinfo {author} {\bibfnamefont {I.}~\bibnamefont {Der{\'e}nyi}},\ and\
  \bibinfo {author} {\bibfnamefont {A.-L.}\ \bibnamefont {Barab{\'a}si}},\
  }\bibfield  {title} {\bibinfo {title} {Reducing vortex density in
  superconductors using the `ratchet effect'},\ }\href@noop {} {\bibfield
  {journal} {\bibinfo  {journal} {Nature}\ }\textbf {\bibinfo {volume} {400}},\
  \bibinfo {pages} {337} (\bibinfo {year} {1999})}\BibitemShut {NoStop}%
\bibitem [{\citenamefont {de~Souza~Silva}\ \emph {et~al.}(2006)\citenamefont
  {de~Souza~Silva}, \citenamefont {Van~de Vondel}, \citenamefont {Zhu},
  \citenamefont {Morelle},\ and\ \citenamefont {Moshchalkov}}]{Ratchet_PRB1}%
  \BibitemOpen
  \bibfield  {author} {\bibinfo {author} {\bibfnamefont {C.~C.}\ \bibnamefont
  {de~Souza~Silva}}, \bibinfo {author} {\bibfnamefont {J.}~\bibnamefont {Van~de
  Vondel}}, \bibinfo {author} {\bibfnamefont {B.~Y.}\ \bibnamefont {Zhu}},
  \bibinfo {author} {\bibfnamefont {M.}~\bibnamefont {Morelle}},\ and\ \bibinfo
  {author} {\bibfnamefont {V.~V.}\ \bibnamefont {Moshchalkov}},\ }\bibfield
  {title} {\bibinfo {title} {Vortex ratchet effects in films with a periodic
  array of antidots},\ }\href {https://doi.org/10.1103/PhysRevB.73.014507}
  {\bibfield  {journal} {\bibinfo  {journal} {Phys. Rev. B}\ }\textbf {\bibinfo
  {volume} {73}},\ \bibinfo {pages} {014507} (\bibinfo {year}
  {2006})}\BibitemShut {NoStop}%
\bibitem [{\citenamefont {Villegas}\ \emph
  {et~al.}(2003{\natexlab{b}})\citenamefont {Villegas}, \citenamefont
  {Gonzalez}, \citenamefont {Montero}, \citenamefont {Schuller},\ and\
  \citenamefont {Vicent}}]{Ratchet_PRB2}%
  \BibitemOpen
  \bibfield  {author} {\bibinfo {author} {\bibfnamefont {J.~E.}\ \bibnamefont
  {Villegas}}, \bibinfo {author} {\bibfnamefont {E.~M.}\ \bibnamefont
  {Gonzalez}}, \bibinfo {author} {\bibfnamefont {M.~I.}\ \bibnamefont
  {Montero}}, \bibinfo {author} {\bibfnamefont {I.~K.}\ \bibnamefont
  {Schuller}},\ and\ \bibinfo {author} {\bibfnamefont {J.~L.}\ \bibnamefont
  {Vicent}},\ }\bibfield  {title} {\bibinfo {title} {Directional vortex motion
  guided by artificially induced mesoscopic potentials},\ }\href
  {https://doi.org/10.1103/PhysRevB.68.224504} {\bibfield  {journal} {\bibinfo
  {journal} {Phys. Rev. B}\ }\textbf {\bibinfo {volume} {68}},\ \bibinfo
  {pages} {224504} (\bibinfo {year} {2003}{\natexlab{b}})}\BibitemShut
  {NoStop}%
\bibitem [{\citenamefont {Zhu}\ \emph {et~al.}(2001)\citenamefont {Zhu},
  \citenamefont {VanLook}, \citenamefont {Moshchalkov}, \citenamefont {Zhao},\
  and\ \citenamefont {Zhao}}]{Zhu}%
  \BibitemOpen
  \bibfield  {author} {\bibinfo {author} {\bibfnamefont {B.~Y.}\ \bibnamefont
  {Zhu}}, \bibinfo {author} {\bibfnamefont {L.}~\bibnamefont {VanLook}},
  \bibinfo {author} {\bibfnamefont {V.~V.}\ \bibnamefont {Moshchalkov}},
  \bibinfo {author} {\bibfnamefont {B.~R.}\ \bibnamefont {Zhao}},\ and\
  \bibinfo {author} {\bibfnamefont {Z.~X.}\ \bibnamefont {Zhao}},\ }\bibfield
  {title} {\bibinfo {title} {Vortex dynamics in regular arrays of asymmetric
  pinning centers},\ }\href {https://doi.org/10.1103/PhysRevB.64.012504}
  {\bibfield  {journal} {\bibinfo  {journal} {Phys. Rev. B}\ }\textbf {\bibinfo
  {volume} {64}},\ \bibinfo {pages} {012504} (\bibinfo {year}
  {2001})}\BibitemShut {NoStop}%
\bibitem [{\citenamefont {Villegas}\ \emph {et~al.}(2005)\citenamefont
  {Villegas}, \citenamefont {Gonzalez}, \citenamefont {Gonzalez}, \citenamefont
  {Anguita},\ and\ \citenamefont {Vicent}}]{Villegas}%
  \BibitemOpen
  \bibfield  {author} {\bibinfo {author} {\bibfnamefont {J.~E.}\ \bibnamefont
  {Villegas}}, \bibinfo {author} {\bibfnamefont {E.~M.}\ \bibnamefont
  {Gonzalez}}, \bibinfo {author} {\bibfnamefont {M.~P.}\ \bibnamefont
  {Gonzalez}}, \bibinfo {author} {\bibfnamefont {J.~V.}\ \bibnamefont
  {Anguita}},\ and\ \bibinfo {author} {\bibfnamefont {J.~L.}\ \bibnamefont
  {Vicent}},\ }\bibfield  {title} {\bibinfo {title} {Experimental ratchet
  effect in superconducting films with periodic arrays of asymmetric
  potentials},\ }\href {https://doi.org/10.1103/PhysRevB.71.024519} {\bibfield
  {journal} {\bibinfo  {journal} {Phys. Rev. B}\ }\textbf {\bibinfo {volume}
  {71}},\ \bibinfo {pages} {024519} (\bibinfo {year} {2005})}\BibitemShut
  {NoStop}%
\bibitem [{\citenamefont {Hoshino}\ \emph {et~al.}(2018)\citenamefont
  {Hoshino}, \citenamefont {Wakatsuki}, \citenamefont {Hamamoto},\ and\
  \citenamefont {Nagaosa}}]{Hoshino}%
  \BibitemOpen
  \bibfield  {author} {\bibinfo {author} {\bibfnamefont {S.}~\bibnamefont
  {Hoshino}}, \bibinfo {author} {\bibfnamefont {R.}~\bibnamefont {Wakatsuki}},
  \bibinfo {author} {\bibfnamefont {K.}~\bibnamefont {Hamamoto}},\ and\
  \bibinfo {author} {\bibfnamefont {N.}~\bibnamefont {Nagaosa}},\ }\bibfield
  {title} {\bibinfo {title} {Nonreciprocal charge transport in two-dimensional
  noncentrosymmetric superconductors},\ }\href
  {https://doi.org/10.1103/PhysRevB.98.054510} {\bibfield  {journal} {\bibinfo
  {journal} {Phys. Rev. B}\ }\textbf {\bibinfo {volume} {98}},\ \bibinfo
  {pages} {054510} (\bibinfo {year} {2018})}\BibitemShut {NoStop}%
\bibitem [{\citenamefont {Itahashi}\ \emph
  {et~al.}(2020{\natexlab{a}})\citenamefont {Itahashi}, \citenamefont {Saito},
  \citenamefont {Ideue}, \citenamefont {Nojima},\ and\ \citenamefont
  {Iwasa}}]{Itahashi}%
  \BibitemOpen
  \bibfield  {author} {\bibinfo {author} {\bibfnamefont {Y.~M.}\ \bibnamefont
  {Itahashi}}, \bibinfo {author} {\bibfnamefont {Y.}~\bibnamefont {Saito}},
  \bibinfo {author} {\bibfnamefont {T.}~\bibnamefont {Ideue}}, \bibinfo
  {author} {\bibfnamefont {T.}~\bibnamefont {Nojima}},\ and\ \bibinfo {author}
  {\bibfnamefont {Y.}~\bibnamefont {Iwasa}},\ }\bibfield  {title} {\bibinfo
  {title} {Quantum and classical ratchet motions of vortices in a
  two-dimensional trigonal superconductor},\ }\href
  {https://doi.org/10.1103/PhysRevResearch.2.023127} {\bibfield  {journal}
  {\bibinfo  {journal} {Phys. Rev. Research}\ }\textbf {\bibinfo {volume}
  {2}},\ \bibinfo {pages} {023127} (\bibinfo {year}
  {2020}{\natexlab{a}})}\BibitemShut {NoStop}%
\bibitem [{\citenamefont {Ideue}\ \emph {et~al.}(2020)\citenamefont {Ideue},
  \citenamefont {Koshikawa}, \citenamefont {Namiki}, \citenamefont {Sasagawa},\
  and\ \citenamefont {Iwasa}}]{Ideue2}%
  \BibitemOpen
  \bibfield  {author} {\bibinfo {author} {\bibfnamefont {T.}~\bibnamefont
  {Ideue}}, \bibinfo {author} {\bibfnamefont {S.}~\bibnamefont {Koshikawa}},
  \bibinfo {author} {\bibfnamefont {H.}~\bibnamefont {Namiki}}, \bibinfo
  {author} {\bibfnamefont {T.}~\bibnamefont {Sasagawa}},\ and\ \bibinfo
  {author} {\bibfnamefont {Y.}~\bibnamefont {Iwasa}},\ }\bibfield  {title}
  {\bibinfo {title} {Giant nonreciprocal magnetotransport in bulk trigonal
  superconductor $\mathrm{PbTa}\mathrm{Se}_2$},\ }\href
  {https://doi.org/10.1103/PhysRevResearch.2.042046} {\bibfield  {journal}
  {\bibinfo  {journal} {Phys. Rev. Research}\ }\textbf {\bibinfo {volume}
  {2}},\ \bibinfo {pages} {042046(R)} (\bibinfo {year} {2020})}\BibitemShut
  {NoStop}%
\bibitem [{\citenamefont {Wakatsuki}\ and\ \citenamefont
  {Nagaosa}(2018)}]{Wakatsuki}%
  \BibitemOpen
  \bibfield  {author} {\bibinfo {author} {\bibfnamefont {R.}~\bibnamefont
  {Wakatsuki}}\ and\ \bibinfo {author} {\bibfnamefont {N.}~\bibnamefont
  {Nagaosa}},\ }\bibfield  {title} {\bibinfo {title} {Nonreciprocal current in
  noncentrosymmetric $\mathrm{R}$ashba superconductors},\ }\href
  {https://doi.org/10.1103/PhysRevLett.121.026601} {\bibfield  {journal}
  {\bibinfo  {journal} {Phys. Rev. Lett.}\ }\textbf {\bibinfo {volume} {121}},\
  \bibinfo {pages} {026601} (\bibinfo {year} {2018})}\BibitemShut {NoStop}%
\bibitem [{\citenamefont {Wakatsuki}\ \emph {et~al.}(2017)\citenamefont
  {Wakatsuki}, \citenamefont {Saito}, \citenamefont {Hoshino}, \citenamefont
  {Itahashi}, \citenamefont {Ideue}, \citenamefont {Ezawa}, \citenamefont
  {Iwasa},\ and\ \citenamefont {Nagaosa}}]{MoS2}%
  \BibitemOpen
  \bibfield  {author} {\bibinfo {author} {\bibfnamefont {R.}~\bibnamefont
  {Wakatsuki}}, \bibinfo {author} {\bibfnamefont {Y.}~\bibnamefont {Saito}},
  \bibinfo {author} {\bibfnamefont {S.}~\bibnamefont {Hoshino}}, \bibinfo
  {author} {\bibfnamefont {Y.~M.}\ \bibnamefont {Itahashi}}, \bibinfo {author}
  {\bibfnamefont {T.}~\bibnamefont {Ideue}}, \bibinfo {author} {\bibfnamefont
  {M.}~\bibnamefont {Ezawa}}, \bibinfo {author} {\bibfnamefont
  {Y.}~\bibnamefont {Iwasa}},\ and\ \bibinfo {author} {\bibfnamefont
  {N.}~\bibnamefont {Nagaosa}},\ }\bibfield  {title} {\bibinfo {title}
  {Nonreciprocal charge transport in noncentrosymmetric superconductors},\
  }\href {https://doi.org/10.1126/sciadv.1602390} {\bibfield  {journal}
  {\bibinfo  {journal} {Science Advances}\ }\textbf {\bibinfo {volume} {3}},\
  \bibinfo {pages} {e1602390} (\bibinfo {year} {2017})}\BibitemShut {NoStop}%
\bibitem [{\citenamefont {Zhang}\ \emph {et~al.}(2020)\citenamefont {Zhang},
  \citenamefont {Xu}, \citenamefont {Zou}, \citenamefont {Ai}, \citenamefont
  {Dong}, \citenamefont {Huang}, \citenamefont {Leng}, \citenamefont {Liu},
  \citenamefont {Zhang}, \citenamefont {Jia}, \citenamefont {Peng},
  \citenamefont {Zhao}, \citenamefont {Yang}, \citenamefont {Li}, \citenamefont
  {Guo}, \citenamefont {Haigh}, \citenamefont {Nagaosa}, \citenamefont {Shen},\
  and\ \citenamefont {Xiu}}]{NbSe2}%
  \BibitemOpen
  \bibfield  {author} {\bibinfo {author} {\bibfnamefont {E.}~\bibnamefont
  {Zhang}}, \bibinfo {author} {\bibfnamefont {X.}~\bibnamefont {Xu}}, \bibinfo
  {author} {\bibfnamefont {Y.-C.}\ \bibnamefont {Zou}}, \bibinfo {author}
  {\bibfnamefont {L.}~\bibnamefont {Ai}}, \bibinfo {author} {\bibfnamefont
  {X.}~\bibnamefont {Dong}}, \bibinfo {author} {\bibfnamefont {C.}~\bibnamefont
  {Huang}}, \bibinfo {author} {\bibfnamefont {P.}~\bibnamefont {Leng}},
  \bibinfo {author} {\bibfnamefont {S.}~\bibnamefont {Liu}}, \bibinfo {author}
  {\bibfnamefont {Y.}~\bibnamefont {Zhang}}, \bibinfo {author} {\bibfnamefont
  {Z.}~\bibnamefont {Jia}}, \bibinfo {author} {\bibfnamefont {X.}~\bibnamefont
  {Peng}}, \bibinfo {author} {\bibfnamefont {M.}~\bibnamefont {Zhao}}, \bibinfo
  {author} {\bibfnamefont {Y.}~\bibnamefont {Yang}}, \bibinfo {author}
  {\bibfnamefont {Z.}~\bibnamefont {Li}}, \bibinfo {author} {\bibfnamefont
  {H.}~\bibnamefont {Guo}}, \bibinfo {author} {\bibfnamefont {S.~J.}\
  \bibnamefont {Haigh}}, \bibinfo {author} {\bibfnamefont {N.}~\bibnamefont
  {Nagaosa}}, \bibinfo {author} {\bibfnamefont {J.}~\bibnamefont {Shen}},\ and\
  \bibinfo {author} {\bibfnamefont {F.}~\bibnamefont {Xiu}},\ }\bibfield
  {title} {\bibinfo {title} {Nonreciprocal superconducting $\mathrm{NbSe}_2$
  antenna},\ }\href {https://doi.org/10.1038/s41467-020-19459-5} {\bibfield
  {journal} {\bibinfo  {journal} {Nature Communications}\ }\textbf {\bibinfo
  {volume} {11}},\ \bibinfo {pages} {5634} (\bibinfo {year}
  {2020})}\BibitemShut {NoStop}%
\bibitem [{\citenamefont {Itahashi}\ \emph {et~al.}(2022)\citenamefont
  {Itahashi}, \citenamefont {Ideue}, \citenamefont {Hoshino}, \citenamefont
  {Goto}, \citenamefont {Namiki}, \citenamefont {Sasagawa},\ and\ \citenamefont
  {Iwasa}}]{Itahashi2}%
  \BibitemOpen
  \bibfield  {author} {\bibinfo {author} {\bibfnamefont {Y.~M.}\ \bibnamefont
  {Itahashi}}, \bibinfo {author} {\bibfnamefont {T.}~\bibnamefont {Ideue}},
  \bibinfo {author} {\bibfnamefont {S.}~\bibnamefont {Hoshino}}, \bibinfo
  {author} {\bibfnamefont {C.}~\bibnamefont {Goto}}, \bibinfo {author}
  {\bibfnamefont {H.}~\bibnamefont {Namiki}}, \bibinfo {author} {\bibfnamefont
  {T.}~\bibnamefont {Sasagawa}},\ and\ \bibinfo {author} {\bibfnamefont
  {Y.}~\bibnamefont {Iwasa}},\ }\bibfield  {title} {\bibinfo {title} {Giant
  second harmonic transport under time-reversal symmetry in a trigonal
  superconductor},\ }\href {https://doi.org/10.1038/s41467-022-29314-4}
  {\bibfield  {journal} {\bibinfo  {journal} {Nature Communications}\ }\textbf
  {\bibinfo {volume} {13}},\ \bibinfo {pages} {1659} (\bibinfo {year}
  {2022})}\BibitemShut {NoStop}%
\bibitem [{\citenamefont {Rikken}\ \emph {et~al.}(2001)\citenamefont {Rikken},
  \citenamefont {Folling},\ and\ \citenamefont {Wyder}}]{Rikken}%
  \BibitemOpen
  \bibfield  {author} {\bibinfo {author} {\bibfnamefont {G.~L. J.~A.}\
  \bibnamefont {Rikken}}, \bibinfo {author} {\bibfnamefont {J.}~\bibnamefont
  {Folling}},\ and\ \bibinfo {author} {\bibfnamefont {P.}~\bibnamefont
  {Wyder}},\ }\bibfield  {title} {\bibinfo {title} {Electrical magnetochiral
  anisotropy},\ }\href {https://doi.org/10.1103/PhysRevLett.87.236602}
  {\bibfield  {journal} {\bibinfo  {journal} {Phys. Rev. Lett.}\ }\textbf
  {\bibinfo {volume} {87}},\ \bibinfo {pages} {236602} (\bibinfo {year}
  {2001})}\BibitemShut {NoStop}%
\bibitem [{\citenamefont {Qi}\ \emph {et~al.}(2016)\citenamefont {Qi},
  \citenamefont {Naumov}, \citenamefont {Ali}, \citenamefont {Rajamathi},
  \citenamefont {Schnelle}, \citenamefont {Barkalov}, \citenamefont {Hanfland},
  \citenamefont {Wu}, \citenamefont {Shekhar}, \citenamefont {Sun},
  \citenamefont {Suss}, \citenamefont {Schmidt}, \citenamefont {Schwarz},
  \citenamefont {Pippel}, \citenamefont {Werner}, \citenamefont {Hillebrand},
  \citenamefont {Forster}, \citenamefont {Kampert}, \citenamefont {Parkin},
  \citenamefont {Cava}, \citenamefont {Felser}, \citenamefont {Yan},\ and\
  \citenamefont {Medvedev}}]{Qi}%
  \BibitemOpen
  \bibfield  {author} {\bibinfo {author} {\bibfnamefont {Y.}~\bibnamefont
  {Qi}}, \bibinfo {author} {\bibfnamefont {P.~G.}\ \bibnamefont {Naumov}},
  \bibinfo {author} {\bibfnamefont {M.~N.}\ \bibnamefont {Ali}}, \bibinfo
  {author} {\bibfnamefont {C.~R.}\ \bibnamefont {Rajamathi}}, \bibinfo {author}
  {\bibfnamefont {W.}~\bibnamefont {Schnelle}}, \bibinfo {author}
  {\bibfnamefont {O.}~\bibnamefont {Barkalov}}, \bibinfo {author}
  {\bibfnamefont {M.}~\bibnamefont {Hanfland}}, \bibinfo {author}
  {\bibfnamefont {S.-C.}\ \bibnamefont {Wu}}, \bibinfo {author} {\bibfnamefont
  {C.}~\bibnamefont {Shekhar}}, \bibinfo {author} {\bibfnamefont
  {Y.}~\bibnamefont {Sun}}, \bibinfo {author} {\bibfnamefont {V.}~\bibnamefont
  {Suss}}, \bibinfo {author} {\bibfnamefont {M.}~\bibnamefont {Schmidt}},
  \bibinfo {author} {\bibfnamefont {U.}~\bibnamefont {Schwarz}}, \bibinfo
  {author} {\bibfnamefont {E.}~\bibnamefont {Pippel}}, \bibinfo {author}
  {\bibfnamefont {P.}~\bibnamefont {Werner}}, \bibinfo {author} {\bibfnamefont
  {R.}~\bibnamefont {Hillebrand}}, \bibinfo {author} {\bibfnamefont
  {T.}~\bibnamefont {Forster}}, \bibinfo {author} {\bibfnamefont
  {E.}~\bibnamefont {Kampert}}, \bibinfo {author} {\bibfnamefont
  {S.}~\bibnamefont {Parkin}}, \bibinfo {author} {\bibfnamefont {R.~J.}\
  \bibnamefont {Cava}}, \bibinfo {author} {\bibfnamefont {C.}~\bibnamefont
  {Felser}}, \bibinfo {author} {\bibfnamefont {B.}~\bibnamefont {Yan}},\ and\
  \bibinfo {author} {\bibfnamefont {S.~A.}\ \bibnamefont {Medvedev}},\
  }\bibfield  {title} {\bibinfo {title} {Superconductivity in $\mathrm{W}$eyl
  semimetal candidate $\mathrm{MoTe}_2$},\ }\href
  {https://doi.org/10.1038/ncomms11038} {\bibfield  {journal} {\bibinfo
  {journal} {Nature Communications}\ }\textbf {\bibinfo {volume} {7}},\
  \bibinfo {pages} {11038} (\bibinfo {year} {2016})}\BibitemShut {NoStop}%
\bibitem [{\citenamefont {Rhodes}\ \emph {et~al.}(2021)\citenamefont {Rhodes},
  \citenamefont {Jindal}, \citenamefont {Yuan}, \citenamefont {Jung},
  \citenamefont {Antony}, \citenamefont {Wang}, \citenamefont {Kim},
  \citenamefont {Chiu}, \citenamefont {Taniguchi}, \citenamefont {Watanabe},
  \citenamefont {Barmak}, \citenamefont {Balicas}, \citenamefont {Dean},
  \citenamefont {Qian}, \citenamefont {Fu}, \citenamefont {Pasupathy},\ and\
  \citenamefont {Hone}}]{Rhodes}%
  \BibitemOpen
  \bibfield  {author} {\bibinfo {author} {\bibfnamefont {D.~A.}\ \bibnamefont
  {Rhodes}}, \bibinfo {author} {\bibfnamefont {A.}~\bibnamefont {Jindal}},
  \bibinfo {author} {\bibfnamefont {N.~F.~Q.}\ \bibnamefont {Yuan}}, \bibinfo
  {author} {\bibfnamefont {Y.}~\bibnamefont {Jung}}, \bibinfo {author}
  {\bibfnamefont {A.}~\bibnamefont {Antony}}, \bibinfo {author} {\bibfnamefont
  {H.}~\bibnamefont {Wang}}, \bibinfo {author} {\bibfnamefont {B.}~\bibnamefont
  {Kim}}, \bibinfo {author} {\bibfnamefont {Y.-c.}\ \bibnamefont {Chiu}},
  \bibinfo {author} {\bibfnamefont {T.}~\bibnamefont {Taniguchi}}, \bibinfo
  {author} {\bibfnamefont {K.}~\bibnamefont {Watanabe}}, \bibinfo {author}
  {\bibfnamefont {K.}~\bibnamefont {Barmak}}, \bibinfo {author} {\bibfnamefont
  {L.}~\bibnamefont {Balicas}}, \bibinfo {author} {\bibfnamefont {C.~R.}\
  \bibnamefont {Dean}}, \bibinfo {author} {\bibfnamefont {X.}~\bibnamefont
  {Qian}}, \bibinfo {author} {\bibfnamefont {L.}~\bibnamefont {Fu}}, \bibinfo
  {author} {\bibfnamefont {A.~N.}\ \bibnamefont {Pasupathy}},\ and\ \bibinfo
  {author} {\bibfnamefont {J.}~\bibnamefont {Hone}},\ }\bibfield  {title}
  {\bibinfo {title} {Enhanced superconductivity in monolayer
  $\mathit{T}_\mathrm{d}$-$\mathrm{MoTe}_2$},\ }\href
  {https://doi.org/10.1021/acs.nanolett.0c04935} {\bibfield  {journal}
  {\bibinfo  {journal} {Nano Letters}\ }\textbf {\bibinfo {volume} {21}},\
  \bibinfo {pages} {2505} (\bibinfo {year} {2021})}\BibitemShut {NoStop}%
\bibitem [{\citenamefont {Gan}\ \emph {et~al.}(2019)\citenamefont {Gan},
  \citenamefont {Cho}, \citenamefont {Li}, \citenamefont {Lyu}, \citenamefont
  {Du}, \citenamefont {Wen},\ and\ \citenamefont {Zhang}}]{MoTe2cn}%
  \BibitemOpen
  \bibfield  {author} {\bibinfo {author} {\bibfnamefont {Y.}~\bibnamefont
  {Gan}}, \bibinfo {author} {\bibfnamefont {C.-W.}\ \bibnamefont {Cho}},
  \bibinfo {author} {\bibfnamefont {A.}~\bibnamefont {Li}}, \bibinfo {author}
  {\bibfnamefont {J.}~\bibnamefont {Lyu}}, \bibinfo {author} {\bibfnamefont
  {X.}~\bibnamefont {Du}}, \bibinfo {author} {\bibfnamefont {J.-S.}\
  \bibnamefont {Wen}},\ and\ \bibinfo {author} {\bibfnamefont {L.-Y.}\
  \bibnamefont {Zhang}},\ }\bibfield  {title} {\bibinfo {title} {Giant
  enhancement of superconductivity in few layers $\mathrm{MoTe}_2$},\ }\href
  {https://doi.org/10.1088/1674-1056/ab457f} {\bibfield  {journal} {\bibinfo
  {journal} {Chinese Physics B}\ }\textbf {\bibinfo {volume} {28}},\ \bibinfo
  {pages} {117401} (\bibinfo {year} {2019})}\BibitemShut {NoStop}%
\bibitem [{\citenamefont {Itahashi}\ \emph
  {et~al.}(2020{\natexlab{b}})\citenamefont {Itahashi}, \citenamefont {Ideue},
  \citenamefont {Saito}, \citenamefont {Shimizu}, \citenamefont {Ouchi},
  \citenamefont {Nojima},\ and\ \citenamefont {Iwasa}}]{SrTiO3}%
  \BibitemOpen
  \bibfield  {author} {\bibinfo {author} {\bibfnamefont {Y.~M.}\ \bibnamefont
  {Itahashi}}, \bibinfo {author} {\bibfnamefont {T.}~\bibnamefont {Ideue}},
  \bibinfo {author} {\bibfnamefont {Y.}~\bibnamefont {Saito}}, \bibinfo
  {author} {\bibfnamefont {S.}~\bibnamefont {Shimizu}}, \bibinfo {author}
  {\bibfnamefont {T.}~\bibnamefont {Ouchi}}, \bibinfo {author} {\bibfnamefont
  {T.}~\bibnamefont {Nojima}},\ and\ \bibinfo {author} {\bibfnamefont
  {Y.}~\bibnamefont {Iwasa}},\ }\bibfield  {title} {\bibinfo {title}
  {Nonreciprocal transport in gate-induced polar superconductor
  $\mathrm{SrTiO}_3$},\ }\href {https://doi.org/10.1126/sciadv.aay9120}
  {\bibfield  {journal} {\bibinfo  {journal} {Science Advances}\ }\textbf
  {\bibinfo {volume} {6}},\ \bibinfo {pages} {eaay9120} (\bibinfo {year}
  {2020}{\natexlab{b}})}\BibitemShut {NoStop}%
\bibitem [{\citenamefont {Masuko}\ \emph {et~al.}(2022)\citenamefont {Masuko},
  \citenamefont {Kawamura}, \citenamefont {Yoshimi}, \citenamefont {Hirayama},
  \citenamefont {Ikeda}, \citenamefont {Watanabe}, \citenamefont {He},
  \citenamefont {Maryenko}, \citenamefont {Tsukazaki}, \citenamefont
  {Takahashi}, \citenamefont {Kawasaki}, \citenamefont {Nagaosa},\ and\
  \citenamefont {Tokura}}]{Masuko}%
  \BibitemOpen
  \bibfield  {author} {\bibinfo {author} {\bibfnamefont {M.}~\bibnamefont
  {Masuko}}, \bibinfo {author} {\bibfnamefont {M.}~\bibnamefont {Kawamura}},
  \bibinfo {author} {\bibfnamefont {R.}~\bibnamefont {Yoshimi}}, \bibinfo
  {author} {\bibfnamefont {M.}~\bibnamefont {Hirayama}}, \bibinfo {author}
  {\bibfnamefont {Y.}~\bibnamefont {Ikeda}}, \bibinfo {author} {\bibfnamefont
  {R.}~\bibnamefont {Watanabe}}, \bibinfo {author} {\bibfnamefont {J.~J.}\
  \bibnamefont {He}}, \bibinfo {author} {\bibfnamefont {D.}~\bibnamefont
  {Maryenko}}, \bibinfo {author} {\bibfnamefont {A.}~\bibnamefont {Tsukazaki}},
  \bibinfo {author} {\bibfnamefont {K.~S.}\ \bibnamefont {Takahashi}}, \bibinfo
  {author} {\bibfnamefont {M.}~\bibnamefont {Kawasaki}}, \bibinfo {author}
  {\bibfnamefont {N.}~\bibnamefont {Nagaosa}},\ and\ \bibinfo {author}
  {\bibfnamefont {Y.}~\bibnamefont {Tokura}},\ }\bibfield  {title} {\bibinfo
  {title} {Nonreciprocal charge transport in topological superconductor
  candidate $\mathrm{Bi}_2\mathrm{Te}_3/\mathrm{PdTe}_2$ heterostructure},\
  }\href {https://doi.org/10.1038/s41535-022-00514-x} {\bibfield  {journal}
  {\bibinfo  {journal} {npj Quantum Materials}\ }\textbf {\bibinfo {volume}
  {7}},\ \bibinfo {pages} {104} (\bibinfo {year} {2022})}\BibitemShut {NoStop}%
\bibitem [{\citenamefont {Tinkham}(2004)}]{Tinkham2}%
  \BibitemOpen
  \bibfield  {author} {\bibinfo {author} {\bibfnamefont {M.}~\bibnamefont
  {Tinkham}},\ }\href@noop {} {\emph {\bibinfo {title} {Introduction to
  Superconductivity: 2nd Ed.}}}\ (\bibinfo  {publisher} {Dover Publications},\
  \bibinfo {year} {2004})\BibitemShut {NoStop}%
\bibitem [{\citenamefont {Kwok}\ \emph {et~al.}(2016)\citenamefont {Kwok},
  \citenamefont {Welp}, \citenamefont {Glatz}, \citenamefont {Koshelev},
  \citenamefont {Kihlstrom},\ and\ \citenamefont {Crabtree}}]{Vortex}%
  \BibitemOpen
  \bibfield  {author} {\bibinfo {author} {\bibfnamefont {W.-K.}\ \bibnamefont
  {Kwok}}, \bibinfo {author} {\bibfnamefont {U.}~\bibnamefont {Welp}}, \bibinfo
  {author} {\bibfnamefont {A.}~\bibnamefont {Glatz}}, \bibinfo {author}
  {\bibfnamefont {A.~E.}\ \bibnamefont {Koshelev}}, \bibinfo {author}
  {\bibfnamefont {K.~J.}\ \bibnamefont {Kihlstrom}},\ and\ \bibinfo {author}
  {\bibfnamefont {G.~W.}\ \bibnamefont {Crabtree}},\ }\bibfield  {title}
  {\bibinfo {title} {Vortices in high-performance high-temperature
  superconductors},\ }\href {https://doi.org/10.1088/0034-4885/79/11/116501}
  {\bibfield  {journal} {\bibinfo  {journal} {Reports on Progress in Physics}\
  }\textbf {\bibinfo {volume} {79}},\ \bibinfo {pages} {116501} (\bibinfo
  {year} {2016})}\BibitemShut {NoStop}%
\bibitem [{\citenamefont {Fente}\ \emph {et~al.}(2018)\citenamefont {Fente},
  \citenamefont {Meier}, \citenamefont {Kong}, \citenamefont {Kogan},
  \citenamefont {Bud'ko}, \citenamefont {Canfield}, \citenamefont
  {Guillam\'on},\ and\ \citenamefont {Suderow}}]{Fente}%
  \BibitemOpen
  \bibfield  {author} {\bibinfo {author} {\bibfnamefont {A.}~\bibnamefont
  {Fente}}, \bibinfo {author} {\bibfnamefont {W.~R.}\ \bibnamefont {Meier}},
  \bibinfo {author} {\bibfnamefont {T.}~\bibnamefont {Kong}}, \bibinfo {author}
  {\bibfnamefont {V.~G.}\ \bibnamefont {Kogan}}, \bibinfo {author}
  {\bibfnamefont {S.~L.}\ \bibnamefont {Bud'ko}}, \bibinfo {author}
  {\bibfnamefont {P.~C.}\ \bibnamefont {Canfield}}, \bibinfo {author}
  {\bibfnamefont {I.}~\bibnamefont {Guillam\'on}},\ and\ \bibinfo {author}
  {\bibfnamefont {H.}~\bibnamefont {Suderow}},\ }\bibfield  {title} {\bibinfo
  {title} {Influence of multiband sign-changing superconductivity on vortex
  cores and vortex pinning in stoichiometric high-$\mathit{T}_c\ \
  \mathrm{CaKFe}_4\mathrm{As}_4$},\ }\href
  {https://doi.org/10.1103/PhysRevB.97.134501} {\bibfield  {journal} {\bibinfo
  {journal} {Phys. Rev. B}\ }\textbf {\bibinfo {volume} {97}},\ \bibinfo
  {pages} {134501} (\bibinfo {year} {2018})}\BibitemShut {NoStop}%
\bibitem [{\citenamefont {Hamamoto}\ \emph {et~al.}(2019)\citenamefont
  {Hamamoto}, \citenamefont {Park}, \citenamefont {Ishizuka},\ and\
  \citenamefont {Nagaosa}}]{Hamamoto}%
  \BibitemOpen
  \bibfield  {author} {\bibinfo {author} {\bibfnamefont {K.}~\bibnamefont
  {Hamamoto}}, \bibinfo {author} {\bibfnamefont {T.}~\bibnamefont {Park}},
  \bibinfo {author} {\bibfnamefont {H.}~\bibnamefont {Ishizuka}},\ and\
  \bibinfo {author} {\bibfnamefont {N.}~\bibnamefont {Nagaosa}},\ }\bibfield
  {title} {\bibinfo {title} {Scaling theory of a quantum ratchet},\ }\href
  {https://doi.org/10.1103/PhysRevB.99.064307} {\bibfield  {journal} {\bibinfo
  {journal} {Phys. Rev. B}\ }\textbf {\bibinfo {volume} {99}},\ \bibinfo
  {pages} {064307} (\bibinfo {year} {2019})}\BibitemShut {NoStop}%
\bibitem [{\citenamefont {Palau}\ \emph {et~al.}(2012)\citenamefont {Palau},
  \citenamefont {Monton}, \citenamefont {Rouco}, \citenamefont {Obradors},\
  and\ \citenamefont {Puig}}]{Palau}%
  \BibitemOpen
  \bibfield  {author} {\bibinfo {author} {\bibfnamefont {A.}~\bibnamefont
  {Palau}}, \bibinfo {author} {\bibfnamefont {C.}~\bibnamefont {Monton}},
  \bibinfo {author} {\bibfnamefont {V.}~\bibnamefont {Rouco}}, \bibinfo
  {author} {\bibfnamefont {X.}~\bibnamefont {Obradors}},\ and\ \bibinfo
  {author} {\bibfnamefont {T.}~\bibnamefont {Puig}},\ }\bibfield  {title}
  {\bibinfo {title} {Guided vortex motion in
  $\mathrm{YBa}_2\mathrm{Cu}_3\mathrm{O}_7$ thin films with collective ratchet
  pinning potentials},\ }\href {https://doi.org/10.1103/PhysRevB.85.012502}
  {\bibfield  {journal} {\bibinfo  {journal} {Phys. Rev. B}\ }\textbf {\bibinfo
  {volume} {85}},\ \bibinfo {pages} {012502} (\bibinfo {year}
  {2012})}\BibitemShut {NoStop}%
\bibitem [{\citenamefont {Morgan}\ and\ \citenamefont
  {Ketterson}(1998)}]{Morgan}%
  \BibitemOpen
  \bibfield  {author} {\bibinfo {author} {\bibfnamefont {D.~J.}\ \bibnamefont
  {Morgan}}\ and\ \bibinfo {author} {\bibfnamefont {J.~B.}\ \bibnamefont
  {Ketterson}},\ }\bibfield  {title} {\bibinfo {title} {Asymmetric flux pinning
  in a regular array of magnetic dipoles},\ }\href
  {https://doi.org/10.1103/PhysRevLett.80.3614} {\bibfield  {journal} {\bibinfo
   {journal} {Phys. Rev. Lett.}\ }\textbf {\bibinfo {volume} {80}},\ \bibinfo
  {pages} {3614} (\bibinfo {year} {1998})}\BibitemShut {NoStop}%
\bibitem [{\citenamefont {Choe}\ \emph {et~al.}(2019)\citenamefont {Choe},
  \citenamefont {Jin}, \citenamefont {Kim}, \citenamefont {Choi}, \citenamefont
  {Jo}, \citenamefont {Oh}, \citenamefont {Park}, \citenamefont {Jin},
  \citenamefont {Koo}, \citenamefont {Min}, \citenamefont {Hong}, \citenamefont
  {Lee}, \citenamefont {Baek},\ and\ \citenamefont {Yoo}}]{LAO}%
  \BibitemOpen
  \bibfield  {author} {\bibinfo {author} {\bibfnamefont {D.}~\bibnamefont
  {Choe}}, \bibinfo {author} {\bibfnamefont {M.-J.}\ \bibnamefont {Jin}},
  \bibinfo {author} {\bibfnamefont {S.-I.}\ \bibnamefont {Kim}}, \bibinfo
  {author} {\bibfnamefont {H.-J.}\ \bibnamefont {Choi}}, \bibinfo {author}
  {\bibfnamefont {J.}~\bibnamefont {Jo}}, \bibinfo {author} {\bibfnamefont
  {I.}~\bibnamefont {Oh}}, \bibinfo {author} {\bibfnamefont {J.}~\bibnamefont
  {Park}}, \bibinfo {author} {\bibfnamefont {H.}~\bibnamefont {Jin}}, \bibinfo
  {author} {\bibfnamefont {H.~C.}\ \bibnamefont {Koo}}, \bibinfo {author}
  {\bibfnamefont {B.-C.}\ \bibnamefont {Min}}, \bibinfo {author} {\bibfnamefont
  {S.}~\bibnamefont {Hong}}, \bibinfo {author} {\bibfnamefont {H.-W.}\
  \bibnamefont {Lee}}, \bibinfo {author} {\bibfnamefont {S.-H.}\ \bibnamefont
  {Baek}},\ and\ \bibinfo {author} {\bibfnamefont {J.-W.}\ \bibnamefont
  {Yoo}},\ }\bibfield  {title} {\bibinfo {title} {Gate-tunable giant
  nonreciprocal charge transport in noncentrosymmetric oxide interfaces},\
  }\href@noop {} {\bibfield  {journal} {\bibinfo  {journal} {Nature
  Communications}\ }\textbf {\bibinfo {volume} {10}},\ \bibinfo {pages} {4510}
  (\bibinfo {year} {2019})}\BibitemShut {NoStop}%
\bibitem [{\citenamefont {Ma}\ \emph {et~al.}(2019)\citenamefont {Ma},
  \citenamefont {Xu}, \citenamefont {Shen}, \citenamefont {MacNeill},
  \citenamefont {Fatemi}, \citenamefont {Chang}, \citenamefont {Mier~Valdivia},
  \citenamefont {Wu}, \citenamefont {Du}, \citenamefont {Hsu}, \citenamefont
  {Fang}, \citenamefont {Gibson}, \citenamefont {Watanabe}, \citenamefont
  {Taniguchi}, \citenamefont {Cava}, \citenamefont {Kaxiras}, \citenamefont
  {Lu}, \citenamefont {Lin}, \citenamefont {Fu}, \citenamefont {Gedik},\ and\
  \citenamefont {Jarillo-Herrero}}]{Ma}%
  \BibitemOpen
  \bibfield  {author} {\bibinfo {author} {\bibfnamefont {Q.}~\bibnamefont
  {Ma}}, \bibinfo {author} {\bibfnamefont {S.-Y.}\ \bibnamefont {Xu}}, \bibinfo
  {author} {\bibfnamefont {H.}~\bibnamefont {Shen}}, \bibinfo {author}
  {\bibfnamefont {D.}~\bibnamefont {MacNeill}}, \bibinfo {author}
  {\bibfnamefont {V.}~\bibnamefont {Fatemi}}, \bibinfo {author} {\bibfnamefont
  {T.-R.}\ \bibnamefont {Chang}}, \bibinfo {author} {\bibfnamefont {A.~M.}\
  \bibnamefont {Mier~Valdivia}}, \bibinfo {author} {\bibfnamefont
  {S.}~\bibnamefont {Wu}}, \bibinfo {author} {\bibfnamefont {Z.}~\bibnamefont
  {Du}}, \bibinfo {author} {\bibfnamefont {C.-H.}\ \bibnamefont {Hsu}},
  \bibinfo {author} {\bibfnamefont {S.}~\bibnamefont {Fang}}, \bibinfo {author}
  {\bibfnamefont {Q.~D.}\ \bibnamefont {Gibson}}, \bibinfo {author}
  {\bibfnamefont {K.}~\bibnamefont {Watanabe}}, \bibinfo {author}
  {\bibfnamefont {T.}~\bibnamefont {Taniguchi}}, \bibinfo {author}
  {\bibfnamefont {R.~J.}\ \bibnamefont {Cava}}, \bibinfo {author}
  {\bibfnamefont {E.}~\bibnamefont {Kaxiras}}, \bibinfo {author} {\bibfnamefont
  {H.-Z.}\ \bibnamefont {Lu}}, \bibinfo {author} {\bibfnamefont
  {H.}~\bibnamefont {Lin}}, \bibinfo {author} {\bibfnamefont {L.}~\bibnamefont
  {Fu}}, \bibinfo {author} {\bibfnamefont {N.}~\bibnamefont {Gedik}},\ and\
  \bibinfo {author} {\bibfnamefont {P.}~\bibnamefont {Jarillo-Herrero}},\
  }\bibfield  {title} {\bibinfo {title} {Observation of the nonlinear
  $\mathrm{H}$all effect under time-reversal-symmetric conditions},\ }\href
  {https://doi.org/10.1038/s41586-018-0807-6} {\bibfield  {journal} {\bibinfo
  {journal} {Nature}\ }\textbf {\bibinfo {volume} {565}},\ \bibinfo {pages}
  {337} (\bibinfo {year} {2019})}\BibitemShut {NoStop}%
\bibitem [{\citenamefont {Fei}\ \emph {et~al.}(2018)\citenamefont {Fei},
  \citenamefont {Zhao}, \citenamefont {Palomaki}, \citenamefont {Sun},
  \citenamefont {Miller}, \citenamefont {Zhao}, \citenamefont {Yan},
  \citenamefont {Xu},\ and\ \citenamefont {Cobden}}]{Ferro}%
  \BibitemOpen
  \bibfield  {author} {\bibinfo {author} {\bibfnamefont {Z.}~\bibnamefont
  {Fei}}, \bibinfo {author} {\bibfnamefont {W.}~\bibnamefont {Zhao}}, \bibinfo
  {author} {\bibfnamefont {T.~A.}\ \bibnamefont {Palomaki}}, \bibinfo {author}
  {\bibfnamefont {B.}~\bibnamefont {Sun}}, \bibinfo {author} {\bibfnamefont
  {M.~K.}\ \bibnamefont {Miller}}, \bibinfo {author} {\bibfnamefont
  {Z.}~\bibnamefont {Zhao}}, \bibinfo {author} {\bibfnamefont {J.}~\bibnamefont
  {Yan}}, \bibinfo {author} {\bibfnamefont {X.}~\bibnamefont {Xu}},\ and\
  \bibinfo {author} {\bibfnamefont {D.~H.}\ \bibnamefont {Cobden}},\ }\bibfield
   {title} {\bibinfo {title} {Ferroelectric switching of a two-dimensional
  metal},\ }\href {https://doi.org/10.1038/s41586-018-0336-3} {\bibfield
  {journal} {\bibinfo  {journal} {Nature}\ }\textbf {\bibinfo {volume} {560}},\
  \bibinfo {pages} {336} (\bibinfo {year} {2018})}\BibitemShut {NoStop}%
\bibitem [{\citenamefont {Olson}\ \emph {et~al.}(2001)\citenamefont {Olson},
  \citenamefont {Reichhardt}, \citenamefont {Jank\'o},\ and\ \citenamefont
  {Nori}}]{Nori}%
  \BibitemOpen
  \bibfield  {author} {\bibinfo {author} {\bibfnamefont {C.~J.}\ \bibnamefont
  {Olson}}, \bibinfo {author} {\bibfnamefont {C.}~\bibnamefont {Reichhardt}},
  \bibinfo {author} {\bibfnamefont {B.}~\bibnamefont {Jank\'o}},\ and\ \bibinfo
  {author} {\bibfnamefont {F.}~\bibnamefont {Nori}},\ }\bibfield  {title}
  {\bibinfo {title} {Collective interaction-driven ratchet for transporting
  flux quanta},\ }\href {https://doi.org/10.1103/PhysRevLett.87.177002}
  {\bibfield  {journal} {\bibinfo  {journal} {Phys. Rev. Lett.}\ }\textbf
  {\bibinfo {volume} {87}},\ \bibinfo {pages} {177002} (\bibinfo {year}
  {2001})}\BibitemShut {NoStop}%
\bibitem [{\citenamefont {Song}\ \emph {et~al.}(2017)\citenamefont {Song},
  \citenamefont {Wang}, \citenamefont {Pan}, \citenamefont {Xu}, \citenamefont
  {Wang}, \citenamefont {Li}, \citenamefont {Song}, \citenamefont {Wan},
  \citenamefont {Ye},\ and\ \citenamefont {Dai}}]{Song2017}%
  \BibitemOpen
  \bibfield  {author} {\bibinfo {author} {\bibfnamefont {Q.}~\bibnamefont
  {Song}}, \bibinfo {author} {\bibfnamefont {H.}~\bibnamefont {Wang}}, \bibinfo
  {author} {\bibfnamefont {X.}~\bibnamefont {Pan}}, \bibinfo {author}
  {\bibfnamefont {X.}~\bibnamefont {Xu}}, \bibinfo {author} {\bibfnamefont
  {Y.}~\bibnamefont {Wang}}, \bibinfo {author} {\bibfnamefont {Y.}~\bibnamefont
  {Li}}, \bibinfo {author} {\bibfnamefont {F.}~\bibnamefont {Song}}, \bibinfo
  {author} {\bibfnamefont {X.}~\bibnamefont {Wan}}, \bibinfo {author}
  {\bibfnamefont {Y.}~\bibnamefont {Ye}},\ and\ \bibinfo {author}
  {\bibfnamefont {L.}~\bibnamefont {Dai}},\ }\bibfield  {title} {\bibinfo
  {title} {Anomalous in-plane anisotropic raman response of monoclinic
  semimetal 1$\mathit{T}$-$\mathrm{MoTe}_2$},\ }\href
  {https://doi.org/10.1038/s41598-017-01874-2} {\bibfield  {journal} {\bibinfo
  {journal} {Scientific Reports}\ }\textbf {\bibinfo {volume} {7}},\ \bibinfo
  {pages} {1758} (\bibinfo {year} {2017})}\BibitemShut {NoStop}%
\bibitem [{\citenamefont {Zhou}\ \emph {et~al.}(2017)\citenamefont {Zhou},
  \citenamefont {Huang}, \citenamefont {Tatsumi}, \citenamefont {Wu},
  \citenamefont {Guo}, \citenamefont {Bie}, \citenamefont {Ueno}, \citenamefont
  {Yang}, \citenamefont {Zhu}, \citenamefont {Kong}, \citenamefont {Saito},\
  and\ \citenamefont {Dresselhaus}}]{Zhou2017}%
  \BibitemOpen
  \bibfield  {author} {\bibinfo {author} {\bibfnamefont {L.}~\bibnamefont
  {Zhou}}, \bibinfo {author} {\bibfnamefont {S.}~\bibnamefont {Huang}},
  \bibinfo {author} {\bibfnamefont {Y.}~\bibnamefont {Tatsumi}}, \bibinfo
  {author} {\bibfnamefont {L.}~\bibnamefont {Wu}}, \bibinfo {author}
  {\bibfnamefont {H.}~\bibnamefont {Guo}}, \bibinfo {author} {\bibfnamefont
  {Y.-Q.}\ \bibnamefont {Bie}}, \bibinfo {author} {\bibfnamefont
  {K.}~\bibnamefont {Ueno}}, \bibinfo {author} {\bibfnamefont {T.}~\bibnamefont
  {Yang}}, \bibinfo {author} {\bibfnamefont {Y.}~\bibnamefont {Zhu}}, \bibinfo
  {author} {\bibfnamefont {J.}~\bibnamefont {Kong}}, \bibinfo {author}
  {\bibfnamefont {R.}~\bibnamefont {Saito}},\ and\ \bibinfo {author}
  {\bibfnamefont {M.}~\bibnamefont {Dresselhaus}},\ }\bibfield  {title}
  {\bibinfo {title} {Sensitive phonon-based probe for structure identification
  of 1$\mathit{T}'$$\mathrm{MoTe}_2$},\ }\href
  {https://doi.org/10.1021/jacs.7b03445} {\bibfield  {journal} {\bibinfo
  {journal} {Journal of the American Chemical Society}\ }\textbf {\bibinfo
  {volume} {139}},\ \bibinfo {pages} {8396} (\bibinfo {year}
  {2017})}\BibitemShut {NoStop}%
\bibitem [{\citenamefont {Saito}\ \emph {et~al.}(2015)\citenamefont {Saito},
  \citenamefont {Kasahara}, \citenamefont {Ye}, \citenamefont {Iwasa},\ and\
  \citenamefont {Nojima}}]{Saito2015}%
  \BibitemOpen
  \bibfield  {author} {\bibinfo {author} {\bibfnamefont {Y.}~\bibnamefont
  {Saito}}, \bibinfo {author} {\bibfnamefont {Y.}~\bibnamefont {Kasahara}},
  \bibinfo {author} {\bibfnamefont {J.}~\bibnamefont {Ye}}, \bibinfo {author}
  {\bibfnamefont {Y.}~\bibnamefont {Iwasa}},\ and\ \bibinfo {author}
  {\bibfnamefont {T.}~\bibnamefont {Nojima}},\ }\bibfield  {title} {\bibinfo
  {title} {Metallic ground state in an ion-gated two-dimensional
  superconductor},\ }\href {https://doi.org/10.1126/science.1259440} {\bibfield
   {journal} {\bibinfo  {journal} {Science}\ }\textbf {\bibinfo {volume}
  {350}},\ \bibinfo {pages} {409} (\bibinfo {year} {2015})}\BibitemShut
  {NoStop}%
\bibitem [{\citenamefont {Li}\ \emph {et~al.}(2020)\citenamefont {Li},
  \citenamefont {Li}, \citenamefont {Zhao},\ and\ \citenamefont {Wu}}]{Li2020}%
  \BibitemOpen
  \bibfield  {author} {\bibinfo {author} {\bibfnamefont {X.-Q.}\ \bibnamefont
  {Li}}, \bibinfo {author} {\bibfnamefont {Z.-L.}\ \bibnamefont {Li}}, \bibinfo
  {author} {\bibfnamefont {J.-J.}\ \bibnamefont {Zhao}},\ and\ \bibinfo
  {author} {\bibfnamefont {X.-S.}\ \bibnamefont {Wu}},\ }\bibfield  {title}
  {\bibinfo {title} {Electrical and thermoelectric study of two-dimensional
  crystal of $\mathrm{NbSe}_2$},\ }\href
  {https://doi.org/10.1088/1674-1056/ab9614} {\bibfield  {journal} {\bibinfo
  {journal} {Chinese Physics B}\ }\textbf {\bibinfo {volume} {29}},\ \bibinfo
  {pages} {087402} (\bibinfo {year} {2020})}\BibitemShut {NoStop}%
\bibitem [{\citenamefont {Behnia}\ and\ \citenamefont
  {Aubin}(2016)}]{Behnia2016}%
  \BibitemOpen
  \bibfield  {author} {\bibinfo {author} {\bibfnamefont {K.}~\bibnamefont
  {Behnia}}\ and\ \bibinfo {author} {\bibfnamefont {H.}~\bibnamefont {Aubin}},\
  }\bibfield  {title} {\bibinfo {title} {Nernst effect in metals and
  superconductors: a review of concepts and experiments},\ }\href
  {https://doi.org/10.1088/0034-4885/79/4/046502} {\bibfield  {journal}
  {\bibinfo  {journal} {Reports on Progress in Physics}\ }\textbf {\bibinfo
  {volume} {79}},\ \bibinfo {pages} {046502} (\bibinfo {year}
  {2016})}\BibitemShut {NoStop}%
\bibitem [{\citenamefont {Vodolazov}\ and\ \citenamefont
  {Peeters}(2005)}]{Vodolazov2005}%
  \BibitemOpen
  \bibfield  {author} {\bibinfo {author} {\bibfnamefont {D.~Y.}\ \bibnamefont
  {Vodolazov}}\ and\ \bibinfo {author} {\bibfnamefont {F.~M.}\ \bibnamefont
  {Peeters}},\ }\bibfield  {title} {\bibinfo {title} {Superconducting rectifier
  based on the asymmetric surface barrier effect},\ }\href
  {https://doi.org/10.1103/PhysRevB.72.172508} {\bibfield  {journal} {\bibinfo
  {journal} {Phys. Rev. B}\ }\textbf {\bibinfo {volume} {72}},\ \bibinfo
  {pages} {172508} (\bibinfo {year} {2005})}\BibitemShut {NoStop}%
\bibitem [{\citenamefont {Cerbu}\ \emph {et~al.}(2013)\citenamefont {Cerbu},
  \citenamefont {Gladilin}, \citenamefont {Cuppens}, \citenamefont {Fritzsche},
  \citenamefont {Tempere}, \citenamefont {Devreese}, \citenamefont
  {Moshchalkov}, \citenamefont {Silhanek},\ and\ \citenamefont
  {de~Vondel}}]{Cerbu2013}%
  \BibitemOpen
  \bibfield  {author} {\bibinfo {author} {\bibfnamefont {D.}~\bibnamefont
  {Cerbu}}, \bibinfo {author} {\bibfnamefont {V.~N.}\ \bibnamefont {Gladilin}},
  \bibinfo {author} {\bibfnamefont {J.}~\bibnamefont {Cuppens}}, \bibinfo
  {author} {\bibfnamefont {J.}~\bibnamefont {Fritzsche}}, \bibinfo {author}
  {\bibfnamefont {J.}~\bibnamefont {Tempere}}, \bibinfo {author} {\bibfnamefont
  {J.~T.}\ \bibnamefont {Devreese}}, \bibinfo {author} {\bibfnamefont {V.~V.}\
  \bibnamefont {Moshchalkov}}, \bibinfo {author} {\bibfnamefont {A.~V.}\
  \bibnamefont {Silhanek}},\ and\ \bibinfo {author} {\bibfnamefont {J.~V.}\
  \bibnamefont {de~Vondel}},\ }\bibfield  {title} {\bibinfo {title} {Vortex
  ratchet induced by controlled edge roughness},\ }\href
  {https://doi.org/10.1088/1367-2630/15/6/063022} {\bibfield  {journal}
  {\bibinfo  {journal} {New Journal of Physics}\ }\textbf {\bibinfo {volume}
  {15}},\ \bibinfo {pages} {063022} (\bibinfo {year} {2013})}\BibitemShut
  {NoStop}%
\bibitem [{\citenamefont {Ji}\ \emph {et~al.}(2016)\citenamefont {Ji},
  \citenamefont {Yuan}, \citenamefont {He}, \citenamefont {Jin}, \citenamefont
  {Zhu}, \citenamefont {Kong}, \citenamefont {Jia}, \citenamefont {Kang},
  \citenamefont {Jin},\ and\ \citenamefont {Wu}}]{Ji2016}%
  \BibitemOpen
  \bibfield  {author} {\bibinfo {author} {\bibfnamefont {J.}~\bibnamefont
  {Ji}}, \bibinfo {author} {\bibfnamefont {J.}~\bibnamefont {Yuan}}, \bibinfo
  {author} {\bibfnamefont {G.}~\bibnamefont {He}}, \bibinfo {author}
  {\bibfnamefont {B.}~\bibnamefont {Jin}}, \bibinfo {author} {\bibfnamefont
  {B.}~\bibnamefont {Zhu}}, \bibinfo {author} {\bibfnamefont {X.}~\bibnamefont
  {Kong}}, \bibinfo {author} {\bibfnamefont {X.}~\bibnamefont {Jia}}, \bibinfo
  {author} {\bibfnamefont {L.}~\bibnamefont {Kang}}, \bibinfo {author}
  {\bibfnamefont {K.}~\bibnamefont {Jin}},\ and\ \bibinfo {author}
  {\bibfnamefont {P.}~\bibnamefont {Wu}},\ }\bibfield  {title} {\bibinfo
  {title} {{Vortex ratchet effects in a superconducting asymmetric ring-shaped
  device}},\ }\href {https://doi.org/10.1063/1.4971835} {\bibfield  {journal}
  {\bibinfo  {journal} {Applied Physics Letters}\ }\textbf {\bibinfo {volume}
  {109}},\ \bibinfo {pages} {242601} (\bibinfo {year} {2016})}\BibitemShut
  {NoStop}%
\bibitem [{\citenamefont {Konczykowski}\ \emph {et~al.}(1991)\citenamefont
  {Konczykowski}, \citenamefont {Burlachkov}, \citenamefont {Yeshurun},\ and\
  \citenamefont {Holtzberg}}]{SB1}%
  \BibitemOpen
  \bibfield  {author} {\bibinfo {author} {\bibfnamefont {M.}~\bibnamefont
  {Konczykowski}}, \bibinfo {author} {\bibfnamefont {L.~I.}\ \bibnamefont
  {Burlachkov}}, \bibinfo {author} {\bibfnamefont {Y.}~\bibnamefont
  {Yeshurun}},\ and\ \bibinfo {author} {\bibfnamefont {F.}~\bibnamefont
  {Holtzberg}},\ }\bibfield  {title} {\bibinfo {title} {Evidence for surface
  barriers and their effect on irreversibility and lower-critical-field
  measurements in $\mathrm{Y}$-$\mathrm{Ba}$-$\mathrm{Cu}$-$\mathrm{O}$
  crystals},\ }\href {https://doi.org/10.1103/PhysRevB.43.13707} {\bibfield
  {journal} {\bibinfo  {journal} {Phys. Rev. B}\ }\textbf {\bibinfo {volume}
  {43}},\ \bibinfo {pages} {13707} (\bibinfo {year} {1991})}\BibitemShut
  {NoStop}%
\bibitem [{\citenamefont {Dewhurst}\ \emph {et~al.}(1996)\citenamefont
  {Dewhurst}, \citenamefont {Cardwell}, \citenamefont {Campbell}, \citenamefont
  {Doyle}, \citenamefont {Balakrishnan},\ and\ \citenamefont {Paul}}]{SB3}%
  \BibitemOpen
  \bibfield  {author} {\bibinfo {author} {\bibfnamefont {C.~D.}\ \bibnamefont
  {Dewhurst}}, \bibinfo {author} {\bibfnamefont {D.~A.}\ \bibnamefont
  {Cardwell}}, \bibinfo {author} {\bibfnamefont {A.~M.}\ \bibnamefont
  {Campbell}}, \bibinfo {author} {\bibfnamefont {R.~A.}\ \bibnamefont {Doyle}},
  \bibinfo {author} {\bibfnamefont {G.}~\bibnamefont {Balakrishnan}},\ and\
  \bibinfo {author} {\bibfnamefont {D.~M.}\ \bibnamefont {Paul}},\ }\bibfield
  {title} {\bibinfo {title} {Determination of the onset of bulk pinning and the
  low-temperature-irreversibility line in $\mathrm{Bi}_2 \mathrm{Sr}_2
  \mathrm{Ca} \mathrm{Cu}_2 \mathrm{O}_{8+\ensuremath{\delta}}$},\ }\href
  {https://doi.org/10.1103/PhysRevB.53.14594} {\bibfield  {journal} {\bibinfo
  {journal} {Phys. Rev. B}\ }\textbf {\bibinfo {volume} {53}},\ \bibinfo
  {pages} {14594} (\bibinfo {year} {1996})}\BibitemShut {NoStop}%
\bibitem [{\citenamefont {Kim}\ \emph {et~al.}(2001)\citenamefont {Kim},
  \citenamefont {Jung}, \citenamefont {Park}, \citenamefont {Lee},
  \citenamefont {Kim}, \citenamefont {Kang},\ and\ \citenamefont {Lee}}]{SB4}%
  \BibitemOpen
  \bibfield  {author} {\bibinfo {author} {\bibfnamefont {M.-S.}\ \bibnamefont
  {Kim}}, \bibinfo {author} {\bibfnamefont {C.~U.}\ \bibnamefont {Jung}},
  \bibinfo {author} {\bibfnamefont {M.-S.}\ \bibnamefont {Park}}, \bibinfo
  {author} {\bibfnamefont {S.~Y.}\ \bibnamefont {Lee}}, \bibinfo {author}
  {\bibfnamefont {K.~H.~P.}\ \bibnamefont {Kim}}, \bibinfo {author}
  {\bibfnamefont {W.~N.}\ \bibnamefont {Kang}},\ and\ \bibinfo {author}
  {\bibfnamefont {S.-I.}\ \bibnamefont {Lee}},\ }\bibfield  {title} {\bibinfo
  {title} {Prominent bulk pinning effect in the $\mathrm{MgB}_2$
  superconductor},\ }\href {https://doi.org/10.1103/PhysRevB.64.012511}
  {\bibfield  {journal} {\bibinfo  {journal} {Phys. Rev. B}\ }\textbf {\bibinfo
  {volume} {64}},\ \bibinfo {pages} {012511} (\bibinfo {year}
  {2001})}\BibitemShut {NoStop}%
\bibitem [{\citenamefont {Zeldov}\ \emph {et~al.}(1994)\citenamefont {Zeldov},
  \citenamefont {Larkin}, \citenamefont {Geshkenbein}, \citenamefont
  {Konczykowski}, \citenamefont {Majer}, \citenamefont {Khaykovich},
  \citenamefont {Vinokur},\ and\ \citenamefont {Shtrikman}}]{Zeldov}%
  \BibitemOpen
  \bibfield  {author} {\bibinfo {author} {\bibfnamefont {E.}~\bibnamefont
  {Zeldov}}, \bibinfo {author} {\bibfnamefont {A.~I.}\ \bibnamefont {Larkin}},
  \bibinfo {author} {\bibfnamefont {V.~B.}\ \bibnamefont {Geshkenbein}},
  \bibinfo {author} {\bibfnamefont {M.}~\bibnamefont {Konczykowski}}, \bibinfo
  {author} {\bibfnamefont {D.}~\bibnamefont {Majer}}, \bibinfo {author}
  {\bibfnamefont {B.}~\bibnamefont {Khaykovich}}, \bibinfo {author}
  {\bibfnamefont {V.~M.}\ \bibnamefont {Vinokur}},\ and\ \bibinfo {author}
  {\bibfnamefont {H.}~\bibnamefont {Shtrikman}},\ }\bibfield  {title} {\bibinfo
  {title} {Geometrical barriers in high-temperature superconductors},\ }\href
  {https://doi.org/10.1103/PhysRevLett.73.1428} {\bibfield  {journal} {\bibinfo
   {journal} {Phys. Rev. Lett.}\ }\textbf {\bibinfo {volume} {73}},\ \bibinfo
  {pages} {1428} (\bibinfo {year} {1994})}\BibitemShut {NoStop}%
\bibitem [{\citenamefont {Kopnin}(2000)}]{Kopnin}%
  \BibitemOpen
  \bibfield  {author} {\bibinfo {author} {\bibfnamefont {N.~B.}\ \bibnamefont
  {Kopnin}},\ }\href@noop {} {\emph {\bibinfo {title} {Theory of Nonequilibrium
  Superconductivity}}}\ (\bibinfo  {publisher} {Oxford University Press, New
  York},\ \bibinfo {year} {2000})\BibitemShut {NoStop}%
\end{thebibliography}%

\end{document}